# Programmed Wrapping and Assembly of Droplets with Mesoscale Polymers

*Dylan M. Barber, Zhefei Yang, Lucas Prévost, Olivia du Roure, Anke Lindner, Todd Emrick,\* and Alfred J. Crosby\**

D. M. Barber, Dr. Z. Yang, Prof. T. Emrick, Prof. A. J. Crosby
Polymer Science and Engineering Department
University of Massachusetts Amherst,
Amherst MA, 01003-9263, USA

L. Prévost, Prof. O. du Roure, Prof. A. Lindner
PMMH, ESPCI Paris, PSL Research University, CNRS
Université de Paris, Sorbonne Université,
Paris, 75005, France

Nature is remarkably adept at using interfaces to build structures, encapsulate reagents, and regulate biological processes. Inspired by Nature, we describe flexible polymer-based ribbons, termed "mesoscale polymers" (MSPs), to modulate interfacial interactions with liquid droplets. This produces unprecedented hybrid assemblies in the forms of flagellum-like structures and MSP-wrapped droplets. Successful preparation of these hybrid structures hinges on interfacial interactions and tailored MSP compositions, such as MSPs with domains possessing distinctly different affinity for fluid-fluid interfaces as well as mechanical properties. *In situ* measurements of MSP-droplet interactions confirm that MSPs possess a negligible bending stiffness, allowing interfacial energy to drive mesoscale assembly. By exploiting these interfacial driving forces, mesoscale polymers are demonstrated as a powerful platform that underpins the preparation of sophisticated hybrid structures in fluids.

## 1. Introduction

Nature provides striking examples of mesoscale assemblies featuring properties and architectures that inspire synthetic replication. Some naturally occurring structures take the form of long, fibrous building blocks that act in concert with spheroids, such as droplets, colloidal particles, or live cells. For example, fiber-in-droplet packing is exemplified by spooling observed in spider capture silk,[1-3] in which a fiber is periodically wetted with aqueous droplets and winds into an internally spooled configuration. The balance between interfacial energy and fiber bending energy drives such assembly, as well as the dissipative, damage-preventing mechanisms activated upon impact-driven disassembly and re-assembly. Another example is the integration of flagella and fimbriae with the membrane of bacteria. These long, flexible mesostructures couple with the vesicle-like core to modulate interfacial interactions with their surroundings.[4-11] These examples illustrate how assemblies of fibers and spheroids with well-controlled interactions and length scales give rise to advantageous properties and performance. While some synthetic systems demonstrate isolated principles of such natural phenomena,[12,13] a robust platform with material-, interfacial-, and geometry-enabled tuning of fiber-spheroid assemblies has yet to be realized.

**Figure 1a** describes our use of polymer ribbons, termed mesoscale polymers (MSPs), at the interface of oil-in-water droplets, in which three modes of interaction were identified: non-adhesion, adhesion without wrapping, and spontaneous wrapping. These interactions are dictated by the critical strain energy release rate, $G_c = \gamma_{ow} + \gamma_{pw} - \gamma_{op}$ (comprising the oil-water, polymer-water, and oil-polymer interfacial tensions), and the critical elasto-adhesive length, $R_c = \sqrt{Et^3/G_c}$, a droplet radius defined by MSP mechanics (Young's modulus $E$), interfacial strength ($G_c$), and geometry (thickness $t$), above which an adhesive MSP spontaneously wraps droplets.[12,13] A pH-responsive trigger embedded in the MSPs controls the observed assembly mode. Figure 1b describes MSPs with segments of alternating compositions, termed mesoscale block copolymers (MSBCPs), such that $G_c$ and $R_c$ are partitioned along the ribbon length. When brought into contact with a droplet of radius $R$, selective wrapping is designed to afford droplets with one or many pendent arms. In this paper, we realize the vision in Figure 1, starting from monomer and copolymer synthesis, fabrication of MS(BC)Ps (thickness $t$ ~ 100-600 nm, width $w$ ~ 10-35 μm, and length 2-4 mm), and MS(BC)P contact with emulsion droplets (radius $R$ = 6-350 μm). Key structures were derived from different ribbon interactions with droplets, including weak adhesion (Figure 1c, far left), spontaneous wrapping (Figure 1c, center left), and selective wrapping by specific MSBCP segments to afford structures with one (Figure 1c center right) or many (Figure 1c far right) arms extending into the surrounding fluid, or a mesoscale micelle. By embedding responsive chemistry into MSPs, we modulate the resulting ribbon/droplet architecture and in turn produce a new materials toolbox of hybrid structures. Moreover, by providing access to a broad array of structures from mesoscale ribbons and droplets, we build a platform of increasingly sophisticated soft materials that begin to emulate the exquisite examples found in Nature.

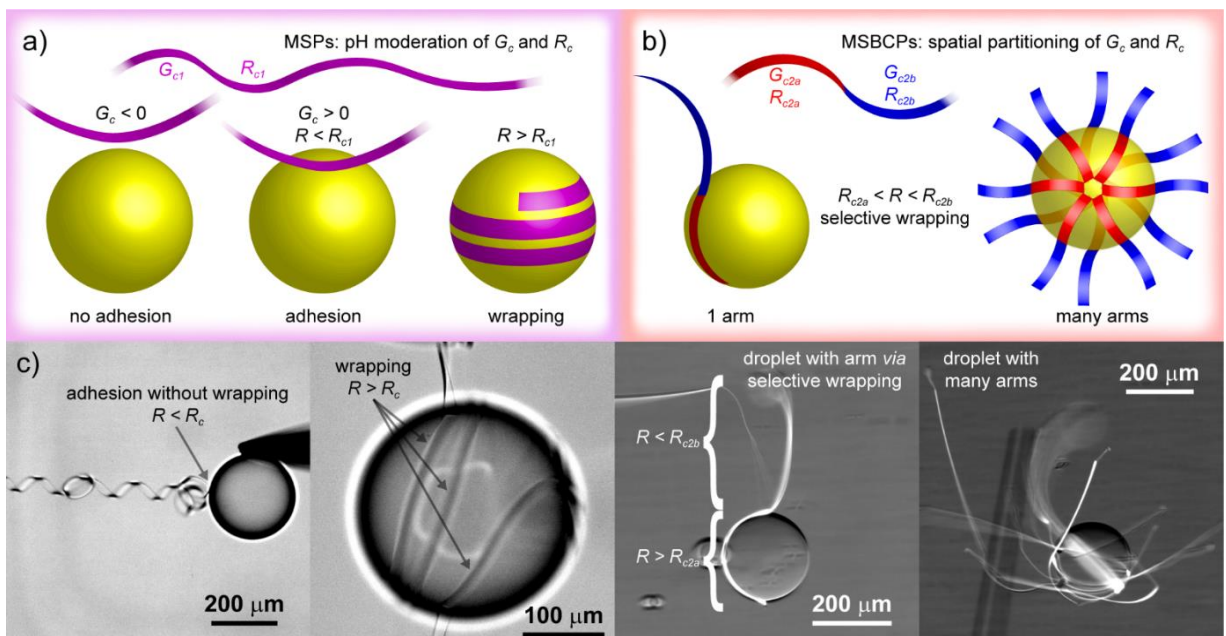

**Figure 1. System design.** MS(BC)P-droplet interactions are dictated by controlling material properties ($G_c$, $E$) and geometry ($t$, $R$) via pH and spatial partitioning: a) MSPs adopt non-adhesive (left), adhesive (center), and wrapped (right) interaction modes, stemming from the pH-dependent work of adhesion ($G_c$) and the relative size of the droplet radius $R$ and critical elasto-adhesive length $R_c$; b) MSBCPs, with segments of alternating composition, $G_c$, and $R_c$, enable selective wrapping for all droplet radii $R_{c2a} < R < R_{c2b}$, affording droplets with 1 (left) or many (right) arms; c) micrographs (left to right) of MSPs in adhesive ($R < R_{c1}$) and wrapped ($R > R_{c1}$) modes, and MSBCPs in selectively wrapped ($R_{c2a} < R < R_{c2b}$) modes with 1 or many arms.

## 2. Materials preparation

The MSPs described in this work were prepared with reactive and functional polymers using flow-coating methods we described previously.[14-16] The polymers were designed to exhibit pH response (polymer **1**) and amenability to photopatterning (polymer **2**), as shown in **Figure 2a**. Polymer **1** ($M_n$ = 38 kDa, Đ = 2.7) was prepared by free radical copolymerization of dimethylaminoethyl methacrylate (DMAEMA) with 5 mole percent of benzophenone methacrylate (BPMA) and 1 mole percent of fluorescein-*o*-methacrylate (FMA). The tertiary amines enable pH response by transitioning from charge neutral to cationic with increasing acidity,[17-20] while BPMA imparts a crosslinking mechanism and FMA contributes fluorescence to aid visualization. Copolymer **2** ($M_n$ = 21 kDa, Đ = 2.2) was prepared by free radical polymerization of *t*-butyl methacrylate (TBMA) with 2 mole percent of glycidyl methacrylate (GMA), 4 mole percent of triphenylsulfonium 4-vinylbenzenesulfonate (TPS4VBS), and 0.2 mole percent of rhodamine B methacrylate (RBMA). In polymer **2**, the aromatic sulfonium sulfonate comonomer functions as a photoacid generator upon UV exposure to trigger acid-catalyzed deprotection of the *t*-butyl esters and crosslinking via the glycidyl ethers, affording MSPs with segments of alternating composition, termed mesoscale block copolymers (MSBCPs).[21]

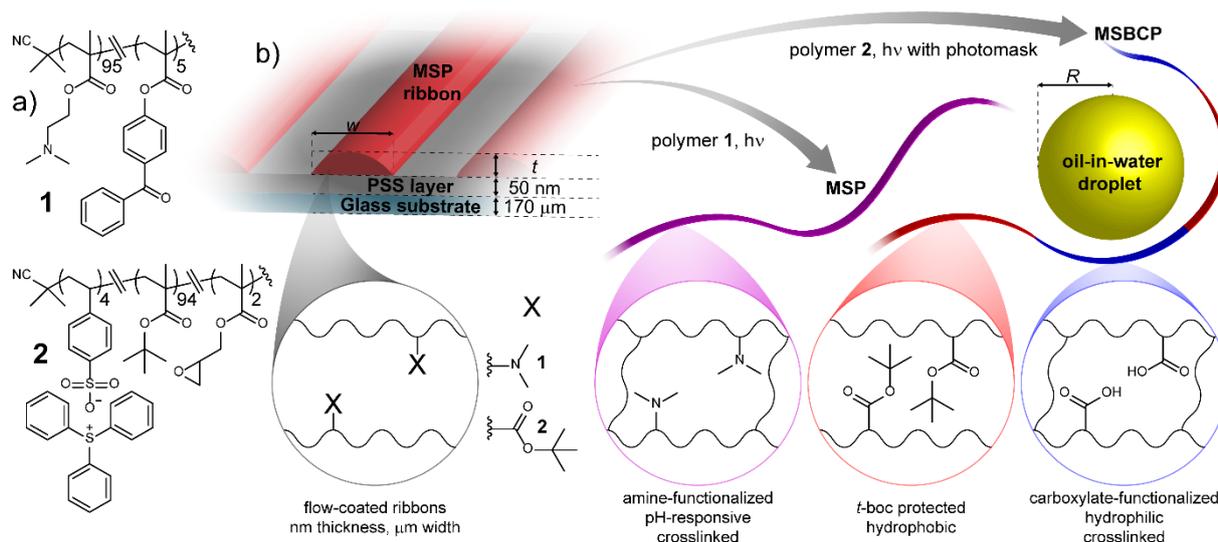

**Fig. 2. Experimental design.** a) Structure of PDMAEMA copolymer **1** and PTBMA copolymer **2** used to prepare ribbons; b) copolymers were flow-coated onto a PSS-coated glass slide to afford ribbons of thickness $t$ and width $w$, with functionality determined by copolymer selection, then irradiated to afford MSPs or MSBCPs (structural representations simplified for clarity).

To prepare the MSPs, a clean glass slide (24 mm x 40 mm x 170 μm) was coated with a ~50 nm layer of poly(styrene sulfonate) (PSS, sodium salt) at 2 or 4 mm intervals to afford stripes of bare glass ~100 μm wide, over which was flow-coated a toluene solution of polymer **1** or **2** (Figure 2b left).[14-16] The substrate was translated in 1 mm intervals at 3 mm s$^{-1}$, with a 1.1-1.5 s delay between steps to deposit the MSPs. The ribbons were then irradiated i) at $\lambda = 365$ nm (3300 mJ cm$^{-2}$) (copolymer **1**) to afford a crosslinked PDMAEMA network (schematic Figure 2b, purple) or at ii) $\lambda = 254$ nm (200-695 mJ cm$^{-2}$) through a photomask, then heated to 150 °C for 60 s (copolymer **2**), to afford MSBCPs with alternating segments of hydrophobic, glassy PTBMA and hydrophilic, crosslinked poly(methacrylic acid) (PMAA, Figure 2b, red and blue, **Figure S1**). The ribbons were cut into 2-4 mm long segments with a CO$_2$ laser engraver ($\lambda = 10.6$ μm) and subjected to reactive ion etching with O$_2$ plasma for 30 s to remove any residual polymer film between the MSPs. The MSPs were released from the substrate by submerging the sample in an aqueous solution to dissolve the underlying PSS layer, then brought into contact with oil-in-water droplets; the resulting assemblies were studied as a function of their interfacial activity ($G_c$) and critical elasto-adhesive length ($R_c$).

## 3. Controlling ribbon-droplet architectures with pH

Experiments with MSPs prepared from copolymer 1 were performed in pH 1-10 buffer solutions using individual perfluorodecalin (PFD) droplets ($R$ = 6-350 μm) to avoid coalescence. Pendent drop tensiometry revealed the oil-water surface tension $\gamma_{ow}$ to be roughly constant (~50 mN m$^{-1}$) across this pH range. Droplet-to-MSP contact was achieved using a glass microcapillary fixed to a hand-controlled micromanipulator (Figure 3a-b). Droplets were introduced by inflation at the microcapillary tip or by emulsification and injection *via* pipette.

The optical micrograph in Figure 3b features an MSP adhered end-on to the surface of a PFD droplet, alongside the microcapillary tip. The schematics in Figure 3c illustrate a typical experimental setup. The microcapillary tube and translating stage are used to probe MSP/droplet interactions by moving droplets through the fluid phase; pH-dependent assembly spans weak adhesion, possibly mediated by non-uniformities on the MSP surface, to spontaneous wrapping. We note that MSPs were observed to spontaneously curve into wavy structures or well-defined helices, especially in aqueous environments from pH 1-6; the observed curvature, a function of MSP mechanical properties and interfacial interactions with the surrounding aqueous phase, was used to estimate a pH-independent copolymer modulus of ~200 MPa by helix extension in viscous flow (details in SI).[15,22,23]

### 3.1. Weak adhesion modes: ribbon stretching and flagellum-like assemblies

From pH 1-6, MSPs and droplets were observed to slide past one another upon contact, with adhesion occurring randomly along the MSP. Figure 3d (left) shows sequential frames from **Video S1**, in which a coiled MSP (helix radius = 38 μm) is suspended between the substrate and an adhered droplet ($R$ = 132 μm). By translating the substrate, the helix transitions from unstretched (top) to extended (center), to fully detached from the droplet (bottom), recoiling like a stretched spring. This adhesion is too weak to macroscopically deform the droplet before detachment. **Video S2** illustrates similar adhesion at pH 4, while **Video S3** displays an example of interfacial slip along a smooth MSP helix at pH 6. At pH 8, the adhesion occurred at the MSP ends (Figure 3c,d center) to afford flagellum-like structures. **Video S4** shows a droplet attached to an MSP segment (length ~400 μm) that is pushed through the fluid with the capillary tip to demonstrate i) adhesion between the droplet and MSP end and ii) a lack of adhesion along the MSP face. This flagellum-like assembly was maintained while the MSP was stretched (Figure 3d center; Video S4), but when the droplet was brought into contact with the MSP face (time $T$ ~ 0.4-0.8 s) the two faces slid past one another without adhering. We speculate that these distinct adhesion modes may result from laser cutting ($CO_2$ laser, $\lambda$ = 10.6 μm) of the MSPs after flow-coating, which heats the material[24] and potentially alters its surface composition (*i.e.*, *via* oxidation), $G_c$, and roughness.[25,26] We note that MSPs that were stored under ambient conditions for ~3 weeks before release into pH 8 buffer qualitatively exhibited a decrease in selectivity for adhesion at the end.

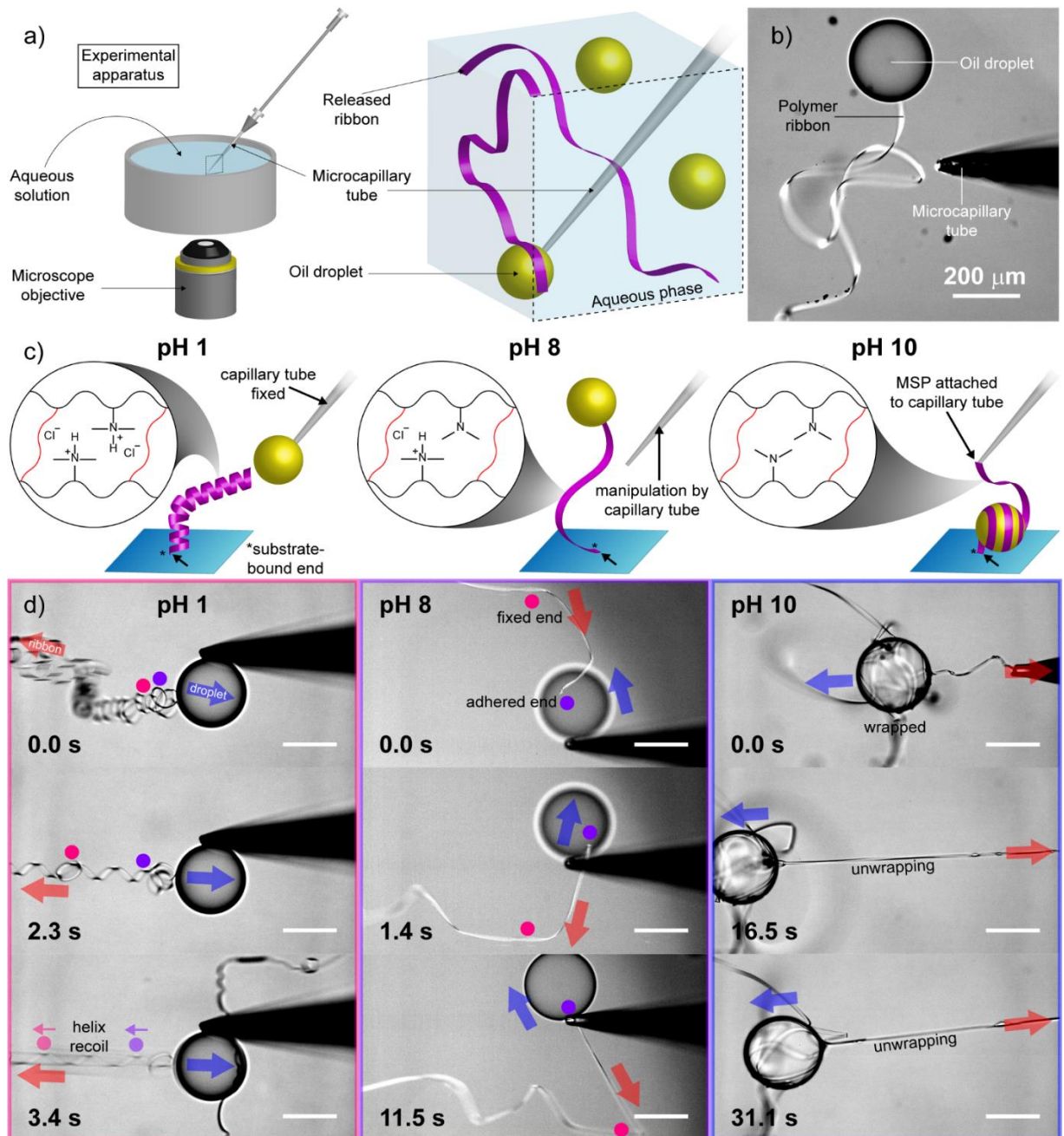

**Figure 3. pH-dependent MSP interfacial activity.** a) Schematic of experimental apparatus: a submerged microcapillary tube was fixed to a micromanipulator and used to move MSPs and droplets through the solution; b) a frame of data featuring a flagellum-like MSP-droplet assembly. c) schematics of experimental design: at pH 1-6 (left), the droplet was fixed to the microcapillary tube and the MSP manipulated *via* translation of the substrate-adhered end; flagellum-like assemblies at pH 8 (center) and spooled assemblies (right) were manipulated by translation of the microcapillary tube and the substrate; d) sequential frames of MSP-droplet assemblies: (left, pH 1) weak, defect-mediated adhesion ($R < R_c$) that detached without macroscopic droplet deformation; (center, pH 8) flagellum-like assembly, and (right, pH 10) unwrapping an assembly where $R > R_c$. Red and blue arrows indicate relative motion of the droplet and the MSP fixed end. Scale bars 200 μm.

## 3.2. Capillary wrapping

At pH 10, the MSPs were observed to spontaneously wrap the droplets upon contact between the ribbon face and fluid-fluid interface, suggesting both large $G_c$ and $R > R_c$. This wrapping event is in stark contrast to the weak adhesion observed at lower pH and marks a transition from polycation (in acidic solution) to neutral polymer (in basic solution, Figure 3c inset structures),[17-20] while a pH-independent $E$ and $\gamma_{ow}$ implicate the polymer surface chemistry as the driving force. Wrapping continued until terminated by one of several mechanisms, including: i) onset of tension in the MSP, supplied by MSP adhesion to the substrate or microcapillary tip; ii) wrapping over an existing coil rather than available oil-water interface; or iii) consumption of the entire MSP length, to afford droplets with partial interfacial coverage. The wrapped droplets were subsequently unwrapped by withdrawing the MSP *via* the microcapillary tube (**Video S5**). Figure 3c (right) schematically depicts the experimental design, while Figure 3d (right) displays frames from Video S5 that show *clean unwinding of millimeters of an MSP while it maintains its structural integrity*. The unwound MSPs then wrap the droplets again when tension is released and the wrapping/unwrapping cycles were repeated up to three times, without noticeable change, for a given MSP-droplet pair. **Video S6** and **S7** demonstrate cases of partial rewrapping to create assemblies in which droplets are decorated with arms that extend into the continuous phase. Because wrapping stops when the MSP wraps upon itself, we infer that it is confined to the oil-water interface, and further, that the wrapping mechanism requires an uninterrupted 3-phase contact line at the wrapping edge.

From a mechanics standpoint, the MSP-wrapped droplets can be described by a thin, wide elastic beam confined to a curved oil-water interface.[12] The components of a wrapped assembly of contact length $L_c$ include bending ($U_b = EI_{yy}L_c/2R^2$) and adhesion ($U_\gamma = G_c w L_c$) energies, where $E$ is the elastic modulus, $I_{yy}$ is the second moment of inertia for axial wrapping, and $G_c$ is critical strain energy release rate. When $R = R_c$, the wrapped and unwrapped states are energetically equivalent, affording $R_c = \sqrt{\frac{EI_{yy}}{2G_c w}} \sim \sqrt{\frac{Et^3}{G_c}}$. Thus, for $R < R_c$ we expect adhesion without wrapping, while for $R > R_c$ we expect spontaneous wrapping. This relationship was studied as a function of droplet radius $R$ in the experiments shown in **Figure 4**. In Figure 4a (and **Video S8**), the microcapillary tip was positioned adjacent to an MSP and used to introduce a droplet, which grew until it contacted the MSP. Figure 4a (left) shows the system at $T = 0.4$ s, immediately before contact and wrapping. To the left, the MSP is fixed to the glass substrate, and to the right, it is unconstrained and free to wrap the droplet. At $T = 11$ seconds (Figure 4a, center), wrapping had nearly advanced one turn around the droplet, and the two wrapping edges passed by one another at $T = 1.4$ s. Approximating wrapping at the droplet circumference, each wrapping edge advanced at ~350 μm s$^{-1}$. After $T = 1.6$ s, the free MSP end was completely wrapped, while the slack between the droplet and the fixed end was pulled tight at $T = 7.6$ s (Figure 4a (right) and final frames of Video S8).

To examine the impact of droplet size on wrapping, a ribbon-wrapped droplet with radius $R = 279$ μm was pierced with the microcapillary tip and oil was continuously withdrawn to reduce the droplet radius (Figure 4b). At $R = 136$ μm, deflation stopped as applied force from the tip translated the droplet without piercing the surface. Despite the decrease in droplet dimensions, the droplet remained wrapped, with an appearance of more substantial interfacial coverage. Even in the presence of small droplets ($R \sim 6$-$30$ μm) prepared by emulsification *via* pipette, wrapping occurred such that MSPs effectively connected multiple droplets in series. For example, Figure 4c shows brightfield (left) and fluorescence (right) micrographs of an MSP ($w = 14$ μm) wrapped around 13 droplets as small as $R = 6$ μm (droplet 7). For even smaller droplets, where $R < w$, we anticipate edgewise wrapping dictated by a lateral moment of inertia $I_{xx}$, which becomes infinitesimally small as MSP thickness tapers toward the edges (**Figure S2**). Accordingly, we expect wrapping even in cases where the thickness $t$ of the MSP central axis might otherwise prohibit lengthwise wrapping.

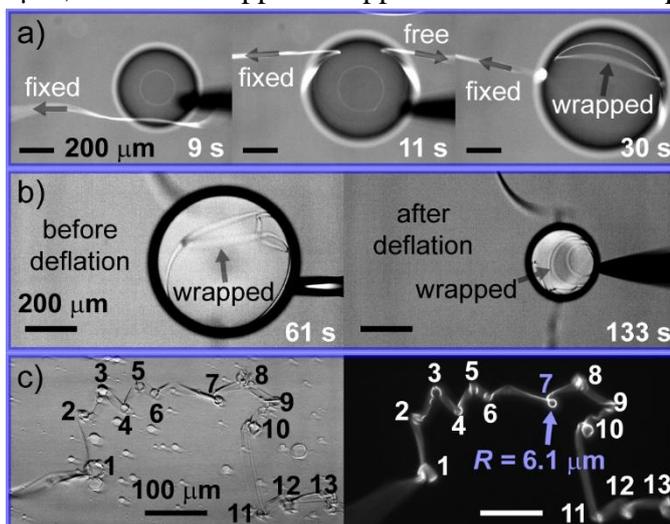

**Figure 4. Critical elastoadhesive dimension $R_{c1}$.** a) time points of a droplet inflated until (left) ribbon contact, (center) mid-wrap, and (right) pulled tight against the substrate-adhered end; b) deflating a pre-wrapped droplet to $R = 136$ μm without any unwrapping; c) bright-field (left) and fluorescence (right) micrographs showing complete wrapping of droplets with diameter $\leq w = 14$ μm.

### 3.3. Evaluating MSP-droplet interactions

The energy landscape of elasto-adhesive MSP wrapping, as described by $G_c$, in pH 10 buffer was probed by measuring the peel force, $F_c$, required to separate a wrapped MSP from the droplet surface. As described in **Figure 5**, these measurements utilized deflection of a single carbon fiber fixed to the end of a glass capillary tube that was dipped into a cyanoacrylate glue and cured to afford a cantilever with a hydrophobic, adhesive bead near the tip. A sample of MSPs was released into the buffer and PFD droplets were introduced by pipette. The cantilever was brought into contact with a PFD droplet *via* a micromanipulator, which adhered to the cured poly(cyanoacrylate) bead, then the cantilever-bound droplet was brought into contact with an MSP to initiate spontaneous wrapping (Figure 5a). For ribbons with one end fixed to the substrate, the MSP-droplet assembly was loaded by substrate translation, enabling direct quantification of the applied force by measuring cantilever deflection. The applied force increased linearly as the MSP stretched and the droplet deformed, as shown by the 3-phase contact line meniscus (Figure 5b), until unwrapping began at a critical force, $F_c$. Fig 5c-d and **Video S9** follow the progress of an experiment with droplet radius $R = 88$ μm through two complete loading cycles, with an unloading step in between the cycles. Force (Fig 5c, left) and the applied energy release rate $G$ (right, describing the energetics of separating the interface) are plotted as a function ribbon length ($L_R$) between its fixed end and the droplet contact point; on the second cycle, the MSP was unwrapped until detachment, when the ribbon contact length was exhausted. The loading curve exhibited two distinct regimes: linear loading, in which the force increased monotonically with the droplet-to-fixed-end MSP length ($L_R$), followed by a plateau of sustained peel at constant force ($F_c$, blue data points) of 2.6 μN;

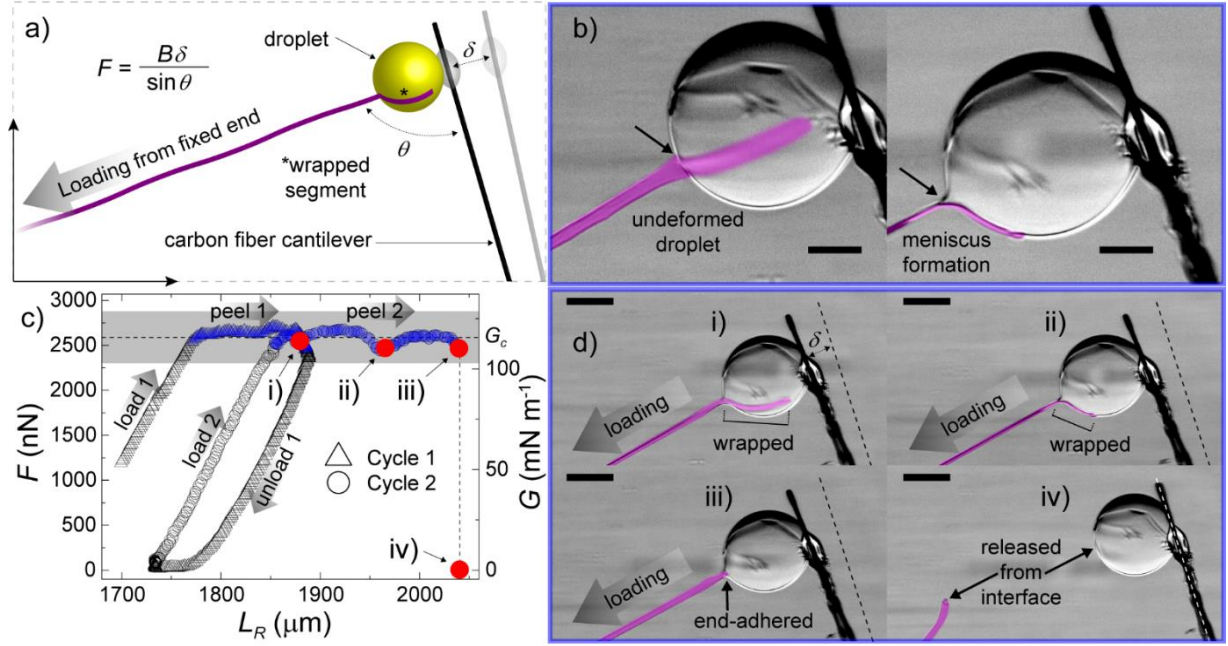

**Figure 5. Measuring MSP peel force and $G_c$ by cantilever deflection**. a) Schematic of experimental apparatus: a droplet is adhered to the tip of a carbon fiber cantilever and partially wrapped by an MSP of suspended length $L_R$. The assembly is loaded by substrate translation to deflect the cantilever by distance $\delta$; b) false color micrographs of the system at low load (left, no MSP-droplet meniscus) and while peeling (right, meniscus formation); c) force-$L_R$ plot of a typical experiment, in which the sample is cycled through two load-peel events, then peeled until rupture. Blue data points denote a visible meniscus and correspond to $F_c$ and $G_c$; cycle averages were combined to estimate peel force and $G_c$ (dashed reference line with 95% confidence in gray); red points i-iv) correspond to frames d) i-iv); d) sequential frames from the same peel experiment. The MSP unwraps (i, ii) from the droplet surface until only the end adheres (iii), then is released from the interface (iv). False coloration highlights the MSP. Scale bars b) 50 μm and d) 100 μm.

the initial loading slope was consistent from cycle to cycle, as was $F_c$. During unloading, the linear force-$L_R$ curve matched the slope of the loading curve, suggesting elastic recovery in the stretched MSP. At $F = 0$, ~50 μm of visible slack spontaneously rewrapped the droplet.

The second load cycle followed a similar stretch-plateau shape and loading continued *until the MSP detached completely from the fluid-fluid interface and dispersed in water*. Figure 5d corresponds to red data points in Figure 5c during the second loading cycle, with wrapped lengths of ~140 μm (i), ~85 μm (ii), and ~0 μm (end-adhered, iii), marking continuous unwrapping before detachment (iv). The critical force for unwrapping is divided by $w$ (~22 μm, measured from video frames) to define a critical energy release rate, $G_c = 116$ mN m$^{-1}$ for the copolymer 1-PFD interface in this solution (Figure 5c, reference line). For an MSP of thickness $t = 300$ nm and modulus 200 MPa, the critical elasto-adhesive dimension for axial wrapping (bending in $I_{yy}$) $R_c$ ~7 µm. We note that $R_c$ is readily decreased by reducing $t$, which is accomplished easily during ribbon fabrication by flow-coating.[14-16]

## 4. Building droplets with arms by photopatterning ribbons

Photopatterned ribbons prepared from copolymer 2 were used to study additional MSP-droplet assembly modes. Here, composition, geometry, and interfacial chemistry are partitioned to afford MSBCPs, reflecting spatial control of $R_c$ such that only pre-determined segments wrap the droplets. Remarkably, only the hydrophobic segments (composed of PTBMA) were observed to wrap PFD droplets, while the hydrophilic PMAA segments exhibited no wrapping tendency, suggesting that for droplet radii $R$ ~ 60-150 µm, $R_{c,PTBMA} < R < R_{c,PMAA}$.

Droplet-ribbon assemblies with appendages extended into the aqueous phase were realized by photochemically programmed wrapping with specific MSP segments, enabled by controlling domain size *via* the photomask and the number of segments *via* laser engraving. **Figure 6a** describes MSBCP assembly consisting of 1 segment each of deprotected PMAA and protected PTBMA (block length 500 µm) with a PFD droplet ($R$ ~ 110 µm) in water; false color (frame 1) highlights the distinct blocks. Upon contact, the hydrophobic PTBMA block wrapped the droplet until reaching the junction point, affording a droplet with a single PMAA arm (Figure 6a frame 2 and **Video S10**). We note that this mechanism of pendent arm formation is distinct from the pH-dependent methods used to prepare extended structures from MSPs of copolymer 1. Subsequent contact with additional MSBCPs decorated the droplet with a second arm (Figure 6a frame 3 and **Video S11**), and up to 10 arms using mixed assembly modes spanning i) selective wrapping, ii) weak adhesion of PMAA domains, and iii) end-on adhesion (Figure 6a, frame 4 and **Video S12**).

Related structures were obtained by using PMAA-PTBMA-PMAA triblock MSBCPs, decorating droplets with two pendent arms per wrapping step. Figure 6b and **Video S13** describe the use of a droplet of $R$ ~150 µm to pick up the ribbons, which are resting on a substrate in 500 mM NaOH solution. The central PTBMA block was 500 µm in length, with shorter blocks of approximate length ~250 µm in PMAA domains. Here, the crosslinked PMAA domains coiled tightly into helices of $R$ ~ 3.5 µm upon release into solution, suggesting swelling-dependent coiling consistent with MSBCP architectures reported previously.[21] In contrast, the hydrophobic PTBMA domains remained straight until contact with a droplet initiated bending. False coloration in frame 1 of Figure 6b highlights the coiled helical end blocks (blue) and rigid core block (red) of an MSBCP immediately before droplet contact and wrapping. Frames 2-4 represent subsequent frames from Video S13 as the droplet is used to remove additional ribbons from the substrate surface by selective wrapping. Wrapping of additional MSBCPs advances until overlap with those present already. Notably, this does not stop the wrapping events as observed for longer, substrate-adhered PDMAEMA MSPs at pH 10; rather, wrapping was seen to continue by pushing the previously wrapped segments across the interface ($T$ ~7.2-14.0 s).

### 4.1. Quantifying MSBCP segment-droplet interactions

The peel force of PTBMA segments at the PFD-water interface was measured by cantilever deflection. MSBCPs of alternating 50 μm blocks were prepared with one end fixed to the substrate surface, released into pH 10 buffer solution, then brought into contact with a cantilever-bound droplet ($R$ ~ 60 μm). Measurements were made by translating the substrate with the adhered MSBCP end, pulling on the droplet, and measuring the deflection of the attached cantilever. The system was taken through two complete load-unload cycles, then loaded until detaching completely from the droplet surface (**Video S14**). Figure 6c represents successive frames from the first cycle in this experiment, including: (i) an unstretched MSBCP; (ii) loading until slack is removed; (iii) hydrogel segment stretching and droplet deformation; and (iv) peeling (false coloration highlights the hydrophobic (red) and hydrogel (blue) domains). The measured force is shown in Figure 6d, revealing continued loading, without peeling, until a critical load of ~ 1460 nN is reached, when the system transitions to a partially peeled state. For each cycle, the average peel force $F_c$ is taken from blue data points, with a typical value of ~1100 nN. Four data points are highlighted as red triangles, corresponding to Figure 6c.i-iv, revealing the load at each stage of the measurement. Relatively little force (~60 nN, ii) is required to straighten the initially curved (i) PMAA domains, which stretch from ~115 μm (low load, ii) to ~160 μm (1470 nN, iii, immediately before and 990 nN, iv, immediately after peel), then elastically recover during unloading, consistent with expectations for a crosslinked hydrogel. Notably, this strain concentration within hydrophilic PMAA gel domains enabled direct measurement of gel modulus $E_{PMAA}$ ~ 2 MPa by tracking the segmental junction points between PMAA and PTBMA domains. By contrast, we estimated $E_{PTBMA}$ on the order of 1 GPa based on the known $T_{g,PTBMA}$ of 116-118 °C,[27,28] *a 500-fold modulus difference achieved simply by photopattern-mediated swelling.*

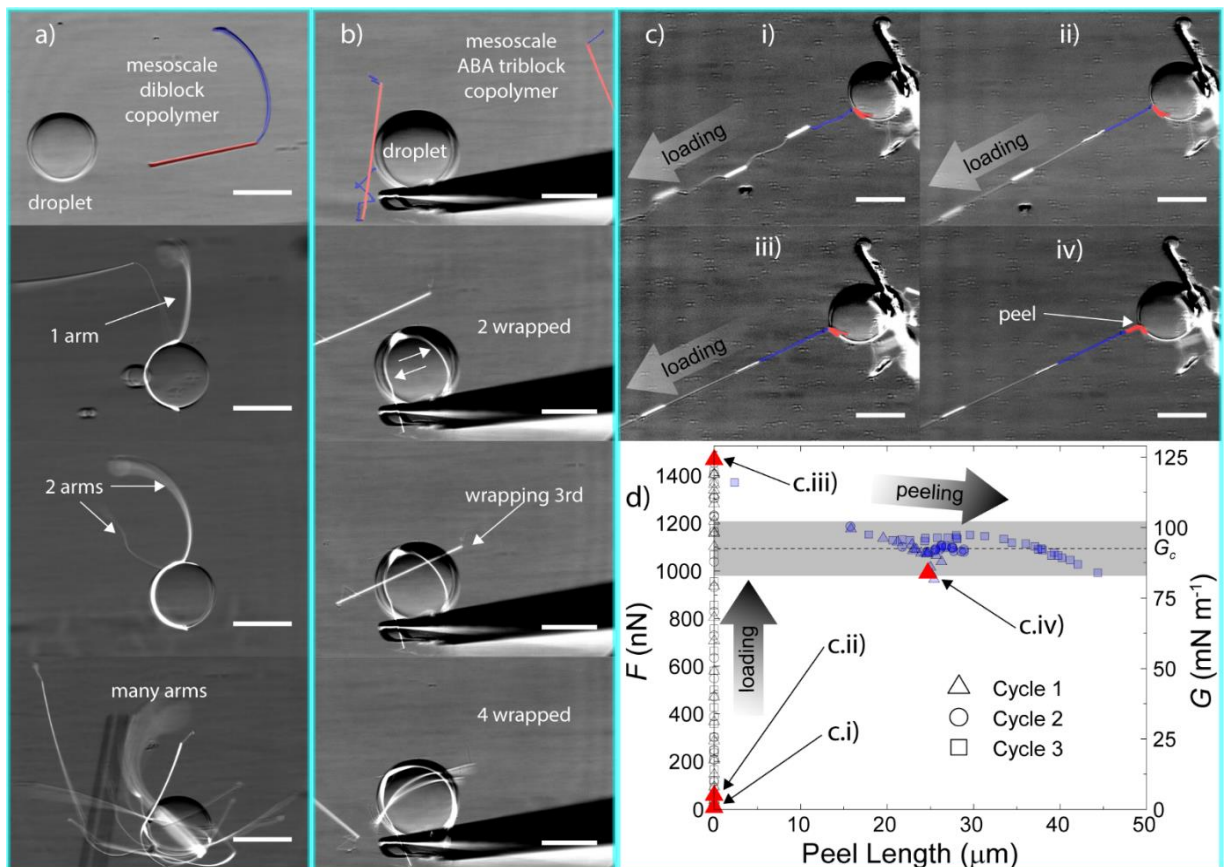

**Figure 6. Droplet-MSBCP assemblies.** Assembly of droplets with a) "diblock" MSP in RO water and b) "triblock" MSP in 500 mM NaOH: a) the droplet and ribbon (false color, top, red = hydrophobic; blue = hydrophilic) are brought into contact; selective wrapping affords a droplet with an arm (2$^{nd}$ frame); subsequent ribbon addition allows installation of 2 (3$^{rd}$ frame) or many (4$^{th}$ frame) arms; b) assembly of MSBCPs (false color, top) enables shorter arms driven to coil in basic solution; c) cantilever deflection of an MSBCP with 1 wrapped segment to quantify peel force; d) plot of measured force $F$ (left axis) and $G$ (right axis) as a function of peel length of an adhered MSBCP subjected to 3 load-peel cycles. Red data points correspond to frames c.i-iv); blue data points denote peeling ($F_c$, $G_c$); cycle averages were combined to calculate peel force and $G_c$ (dashed reference line with 95% confidence in gray). Scale bars a-b) 200 μm and c) 100 μm.
.

The measured peel force ($F_c$) represents the energy per unit length required to unwrap ribbons from the curved oil-water interface. Having demonstrated that capillary interactions dominate bending stiffness at the selected length scales in PDMAEMA MSPs of modulus 200 MPa ($G_c$ = 116 mN m$^{-1}$, $R_c$ = 7 μm for $t$ = 300 nm), we applied the same assumption when measuring MSBCP segments. Dividing $F_c$ by segment width $w$ = 12 μm (measured *via* optical profilometry before release), $G_c$ ~ 93 mN m$^{-1}$ (Figure 6d reference line) was calculated. Thus, for a hydrophobic MSBCP segment with $t$ = 300 nm, $R_c$ = 17 μm, while smaller values are readily accessible by printing thinner MSBCPs. Notably, $G_c$ for MSPs (~116 mN m$^{-1}$) and hydrophobic MSBCP segments (~93 mN m$^{-1}$) are comparable to the oil-water interfacial tension $\gamma_{ow}$ = 51 mN m$^{-1}$ measured by pendent drop tensiometry; moreover, MSBCP adhesion at the oil-water interface ceased upon the addition of a polymer surfactant, further connecting the high energy oil-water and polymer-water interfaces to adhesion and wrapping phenomena. Together, our measurements of $G_c$ and $\gamma_{ow}$ combined with loss of adhesion in the presence of surfactant implicate the oil-water and polymer-water interfaces as a primary driving force for large scale assembly of mesoscale ribbons. Notably, despite a modulus approximately 3 orders of magnitude smaller than the glassy PTBMA domains, the PMAA gel segments *adhered to droplets without wrapping*, suggesting an equally dramatic change in $G_c$ from segment to segment. Thus, MSBCPs possess partitioned domains of alternating physical and mechanical properties, including a 500-fold difference in elastic modulus, and dramatic differences in $G_c$ and $R_c$ that enable selective wrapping and assembly upon contact with oil-in-water droplets.

**5. Conclusion**

In summary, we described the use of compliant, surface-active, mesoscale polymer ribbons to build assemblies with liquid droplets *via* the fluid-fluid interface of the droplets. We adapted a model of cylindrical filaments at droplet surfaces to describe the uniquely flat geometry of MSPs in contact with an oil-in-water droplet, spanning wrapping and non-wrapping interaction modes as a function of a modulus-, geometry-, and $G_c$-dependent elasto-adhesive dimension $R_c$. Using photocrosslinked MSPs derived from copolymer 1, we mapped pH-dependent interactions, ranging from i) weak adhesion ($R_c > R$) from pH 1-8, including flagellum-like architectures formed by selective adhesion at the MSP tip, to ii) spontaneous wrapping at pH 10, producing spools amenable to unwrapping, re-wrapping, and addition of pendent arms. We employed the "built-in" photoacid generators in copolymer 2 to effect chemically amplified deprotection and crosslinking, using a photomask to partition distinct properties into segments along the ribbon length. Within the resulting MSBCP structures, hydrophobic PTBMA segments were observed to selectively wrap oil-in-water droplets independent of pH, enabling the construction of droplets with 1, 2, or many arms extended into solution. Moreover, quantification of $G_c$ and thickness-dependent $R_c$ confirms that the bending compliance and strong interfacial activity of MSPs and MSBCPs affords elasto-adhesive lengths of microns or smaller. Together, these pH-, light-, and spatially-programmable structures provide a robust platform to transform simple soft materials building blocks and interaction modes into sophisticated meso-to-macroscale bio-inspired assemblies.

**Supporting Information**

Supporting Information is available from the Wiley Online Library or from the author.

**Acknowledgements**

This project was supported by the Department of Energy, Office of Basic Energy Sciences, Division of Materials Science and Engineering under award number DE-SC0008876 and a National Defense Science and Engineering Graduate (NDSEG) Fellowship awarded to D.M.B. A.L., L.P. and D.M.B. acknowledge funding from the ERC Consolidator Grant PaDyFlow (Grant Agreement no. 682367).

# Supporting Information

**Programmed Wrapping and Assembly of Droplets with Mesoscale Polymers**

*Dylan M. Barber, Zhefei Yang, Lucas Prévost, Olivia du Roure, Anke Lindner, Todd Emrick,\* and Alfred J. Crosby\**

Methods

*Chemicals.* Methacryloyl chloride, rhodamine B, 4-dimethylaminopyridine (DMAP), N,N'-dicyclohexylcarbodiimide (DCC), 2-hydroxyethyl methacrylate (HEMA), triphenylsulfonium chloride (TPSCl), fluorescein-*O*-methacrylate (FOMA), toluene, perfluorodecalin (PFD), buffer solutions, basic alumina, lithium chloride (LiCl), poly(sodium 4-styrenesulfonate) (PSS, MW 70 kDa, Aldrich), 4-hydroxybenzophenone (4HBP, TCI America), methanol (MeOH), dimethylformamide (DMF), hexanes, isopropanol (IPA, Fisher Scientific), sodium 4-vinylbenzenesulfonate (Na4VBS, Alfa Aesar), and silica gel (Sorbent Technologies) were used as received without further purification. Triethylamine (TEA, Aldrich) and dichloromethane (DCM, Fisher Scientific) were dried over calcium hydride and distilled. 2.1% aqueous ammonium hydroxide solution was prepared by diluting 28 wt% ammonium hydroxide solution (Aldrich) into stirring RO water. 100 mM HCl solution was prepared by dropwise addition of 12.1 N HCl (Fisher Scientific) to a beaker of stirring RO water. 2-(Dimethylamino)ethyl methacrylate (DMAEMA), *tert*-butyl methacrylate (TBMA), and glycidyl methacrylate (GMA, Aldrich) were purified by passage through a plug of basic alumina. 2,2'-Azobisisobutyronitrile (AIBN, Aldrich) was recrystallized from MeOH. Tetrahydrofuran (THF, Fisher Scientific) was dried over sodium benzophenone ketyl, then distilled. $N_2$ gas was dried by passing through Drierite (W.A. Hammond Drierite Company).

*Instrumentation.* $^1$H NMR (500 MHz) spectroscopic data was collected using a Bruker Ascend TM500 spectrometer with a Prodigy cryoprobe. Copolymer molecular weight was estimated against PMMA standards by gel permeation chromatography (GPC), eluting in a mobile phase of 0.01 M LiCl in DMF at 1 mL min$^{-1}$ flow rate (Agilent 1260 Infinity isocratic pump) through a 50 × 7.5 mm PL gel mixed guard column, a 300 × 7.5 mm PL gel 5 μm mixed C column, and a 300 × 7.5 mm PL gel 5 μm mixed D column at 50 °C. Solute was detected using an Agilent 1260 Infinity refractive index detector. UV-ozone (UVO) surface treatment was conducted with a Jelight Company, Inc. Model 342 UVO-Cleaner®. Laser engraving was carried out using a Universal Laser Systems VLS3.50 laser engraver equipped with a 30W $CO_2$ (10.6 μm) laser with 0.005" z-axis offset, 2% power, 40% speed, and 1000 ppi pulse rate. Flow-coating was carried out using a SmarAct, Inc SLC-1780s linear actuator. 365 nm UV-irradiation was performed on a Newport 97435 lamp housing with a Newport 69910 power supply and Newport 6285 Mercury arc lamp or a Suss Micro Tec MA6 Mask Aligner. An OAI Instruments 1000 Watt DUV Exposure System equipped with a DUV 1000 lamp (Advanced Radiation

Corporation) was used for all 254 nm UV irradiation. Reactive Ion Etch (RIE) experiments employed an Advanced Vacuum Vision 320 MkII Reactive Ion Etch System with 50 sccm $O_2(g)$ flow rate, 50 mTorr chamber pressure, 100 W RF power, and 13.56 MHz RF frequency. Microscopy was conducted on an Axio Observer 7 Materials microscope equipped with a Hamamatsu C11440 Orca-Flash4.0 Digital Camera, 2 Eppendorf TransferMan 4r micromanipulators, an X-Cite 120LED (Excelitas Technologies), and Zeiss filter set 38 HE (green fluorescence, copolymer **1**) or 45 (red fluorescence, copolymer **2**). Fourier-transform Infrared (FT-IR) data were collected in attenuated total reflectance mode using a PerkinElmer Spectrum One FT-IR Spectrometer equipped with a Universal ATR Sampling Accessory. Optical profilometry data was collected using a Zygo NewView 7300 Optical Surface Profiler (Amherst) or a Veeko Instruments Wyko NT9100 (Paris). Microcapillary tubes were prepared by drawing glass capillary tubes (ChemGlass, 1.0-1.1 mm O.D.) in a P-1000 Flaming/Brown™ Micropipette Puller System (Sutter Instrument) and the melted ends were opened using an MF-830 Microforge (Narishige International).

*Synthesis of benzophenone methacrylate (BPMA) monomer*. BPMA synthesis was adapted from a reported procedure.[1] In brief, a 500 mL round-bottom flask with a stir bar was flame-dried and purged with dry nitrogen, then 4-hydroxybenzophenone (5.1 g, 25.7 mmol, 1 equivalent) was added against a positive flow of dry $N_2(g)$. The flask was sealed with a septum. Dry TEA (8 mL, 57.4 mmol, 2 equivalents) and dry DCM (75 mL) were added by syringe against positive $N_2(g)$ pressure; the solution was stirred until homogeneous then cooled to 0 °C. Methacryloyl chloride (4.8 mL, 49.6 mmol, 1.93 equivalents) in dry DCM (25 mL) was added dropwise while stirring. The solution was allowed to return to 20 °C where it was stirred for 15.5 h, then concentrated under vacuum, redissolved in ether, and washed with 2.1% aqueous ammonium hydroxide solution. The product was purified by column chromatography (basic alumina as stationary phase, 90:10 hexanes:ethyl acetate as eluent) to afford the desired product as white crystals (4.0 g, 58% yield). $^1$H NMR (500 MHz, $CDCl_3$, δ) 7.82-7.77 (m, 2H, aromatic), 7.75-7.70 (d, 2H, aromatic, J = 7.02 Hz), 7.54-7.49 (t, 1H, aromatic, J = 7.43 Hz), 7.45-7.38 (t, 2H, aromatic, J = 7.68 Hz), 7.21-7.15 (m, 2H, aromatic), 6.34-6.29 (s, 1H, vinyl), 5.75-5.71 (t, 1H, vinyl J = 1.39 Hz), 2.03-1.99 (s, 3H, $CCH_3$).

*Synthesis of rhodamine B methacrylate (RBMA) monomer*. The RBMA synthesis was also adapted from a reported procedure.[2,3] In brief, a 2-neck, 250 mL round-bottom flask with stir bar was flame-dried and purged with dry nitrogen gas, then rhodamine B (10 g, 20.9 mmol, 1 equivalent), DMAP (150 mg, 1.23 mmol, 0.06 equivalents), and DCC (5.2 g, 25.2 mmol, 1.21 equivalents) were added against positive flow of dry $N_2(g)$. The flask was sealed with a septum, then dry DCM (105 mL) and HEMA (3.1 mL, 25 mmol, 1.20 equivalents) were added by syringe. The solution was stirred at 20 °C for 25 h, then concentrated under reduced pressure and purified by column chromatography (silica gel stationary phase, 90:10 DCM:MeOH eluent) and dried under high vacuum to afford a dark purple powder (6.15 g, 50% yield). $^1$H NMR (500 MHz, CDCl3, δ) 8.33-8.26 (d, 1H, aromatic, J = 7.90 Hz), 7.88-7.81 (t, 1H, aromatic, J = 7.45 Hz), 7.79-7.72 (t, 1H, J = 7.68 Hz), 7.35-7.30 (d, 1H, J = 7.50 Hz), 7.10-7.03 (d, 2H, J = 9.45), 6.97-6.90 (dd, 2H, J1 = 9.45 Hz, J2 = 2.25 Hz), 6.82-6.77 (d, 2H, J = 2.20 Hz), 6.05-5.98 (s, 1H, vinyl), 5.58-5.52 (s, 1H, vinyl), 4.33-4.28 (t, OCH2CH2O, J = 4.95 Hz), 4.21-4.16 (t, 2H OCH2CH2O, J = 4.68 Hz), 3.70-3.63 (8H, q, NCH2CH3, J = 7.20 Hz), 1.90-1.85 (s, 3H, methacrylate CCH3), 1.37-1.29 (t, 12H, NCH2CH3, J = 7.05 Hz)

*Synthesis of triphenylsulfonium 4-vinylbenzenesulfonate (TPS-4-VBS) monomer.* TPS-4-VBS was synthesized by adapting a procedure from a literature report.[4] In brief, 94% TPSCl (1.06 g, 3.33 mmol, 1 equivalent) and 90% Na4VBS (767 mg, 3.35 mmol, 1 equivalent) were combined and shaken with 3.3 mL RO water in a 20 mL scintillation vial to afford a brown emulsion. The brown organic phase was removed, and the aqueous phase extracted with 6 x 1 mL DCM. The combined organic phase was diluted to 12 mL, washed with 4 x 1 mL RO water, filtered to remove residual brown solid, concentrated, then diluted with hexanes (1 mL) to induce crystallization. Residual solvent was removed under reduced pressure to afford the desired product as white crystals (1.24 g, 83 % yield). $^1$H NMR: (500 MHz, CDCl$_3$, δ): 7.86-7.81 (d, 2H, 4-vinylbenzenesulfonate aromatic, J = 8.23 Hz), 7.76-7.72 (d, 6H, S$^+$(C$_6$H$_5$)$_3$, J = 7.51 Hz), 7.70-7.66 (t, 3H, S$^+$(C$_6$H$_5$)$_3$, J = 7.42 Hz), 7.64-7.59 (t, 6H, S$^+$(C$_6$H$_5$)$_3$, J = 7.62 Hz), 7.30-7.27 (d, 2H, 4-vinylbenzenesulfonate aromatic, J = 8.20 Hz), 6.70-6.60 (dd, 1H, 4-vinylbenzenesulfonate vinyl, J = 10.89, 17.61 Hz), 5.74-5.65 (d, 1H, 4-vinylbenzenesulfonate vinyl, J = 17.61 Hz), 5.24-5.17 (d, 1H, 4-vinylbenzenesulfonate vinyl, J = 10.97 Hz).

*Synthesis of copolymer 1.* DMAEMA (3.2 mL, 19 mmol, 197 equivalents), BPMA (289 mg, 1.1 mmol, 11 equivalents), FOMA (83 mg, 0.21 mmol, 2 equivalents), and AIBN (15.8 mg, 0.10 mmol, 1 equivalent) were dissolved in a mixture of THF (9 mL) and DMF (1 mL) in a 20 mL scintillation vial containing a stir bar. The vial was sealed with a rubber septum, then degassed with dry N$_2$(g) for 30 minutes while stirring at 20 °C. After removing needles, the septum was covered with a piece of electrical tape and the vial transferred to an aluminum heating block, where the mixture was stirred for 22 hours at 60 °C. The reaction mixture was then precipitated three times in stirring hexanes at 20 °C and dried under vacuum at 60 °C for 18 h to afford the desired product (1.22 g, 36 % yield). $^1$H NMR: (500.13 MHz, CDCl$_3$, δ): 8.07-7.97 (br s, aromatic), 7.89-7.70 (br m, 4H, BPMA aromatic), 7.63-7.54 (br m, 1H, BPMA aromatic), 7.52-7.41 (br m, 2H, BPMA aromatic), 7.30-7.16 (br m, 2H, BPMA aromatic), 6.85-6.42 (br m, FOMA aromatic), 4.25-3.85 (br m, 2H, DMAEMA OC$\underline{H}_2$CH$_2$N), 2.70-2.45 (br m, 2H, DMAEMA OCH$_2$C$\underline{H}_2$N), 2.44-2.11 (br m, 6H, N(CH$_3$)$_2$, 2.11-0.73 (br m, aliphatic backbone, CH$_2$CCH$_3$). $^{13}$C NMR: (125.76 MHz, CDCl$_3$, δ): 195.85-195.12 (s, 1C, BPMA ketone), 178.63-173.61 (br m, ester carbonyl), 154.45-153.66 (m, 1C, BPMA aromatic), 137.86-137.22 (m, 1C, BPMA aromatic), 135.43-134.83 (m, 1C, BPMA aromatic), 132.81-

132.43 (m, 1C, BPMA aromatic), 131.98-131.46 (m, 2C, BPMA aromatic), 130.26-129.85 (s, 2C, BPMA aromatic), 128.69-128.28 (s, 2C, BPMA aromatic), 121.57-121.00 (m, 2C, BPMA aromatic), 63.77-62.55 (m, 1C, DMAEMA OCH$_2$), 57.63-56.98 (m, 1C, DMAEMA CH$_2$N), 55.40-51.52 (br m, 1C, backbone methylene), 46.30-45.47 (s, 2C, DMAEMA N(CH$_3$)$_2$), 45.47-44.49 (br m, 1C, backbone quaternary), 19.42-15.94 (br m, 1C, backbone CH$_3$). GPC: (DMF with 10 mM LiBr, PMMA standards): $M_n$ = 38 kDa, $M_w$ = 104 kDa, Đ = 2.70.

*Synthesis of copolymer 2*. TBMA (2.3 mL, 14 mmol, 89 equivalents), TPS-4-VBS (200 mg, 0.45 mmol, 2.8 equivalents), GMA (39 μL, 0.29 mmol, 1.9 equivalents), RBMA (20 mg, 34 μmol, 0.2 equivalents), and AIBN (26 mg, 0.16 mmol, 1 equivalent) were dissolved in DMF (5 mL) in a 20 mL scintillation vial equipped with a stir bar, then degassed by bubbling for 30 minutes with dry N$_2$(g) while stirring at 20 °C. After degassing, the septum was covered with a piece of electrical tape and the vial was transferred to an aluminum block, where the mixture was stirred at 80 °C for 22 h. The reaction was stopped by cooling to -20 °C, then purified by precipitating into 65:35 water:MeOH, re-dissolving in THF, precipitating three times in stirring hexanes, and finally drying under high vacuum at 20 °C for 18 h to yield the desired product. (1.03 g, 45%). $^1$H NMR: (500.13 MHz, CDCl$_3$, δ): 7.88-7.80 (d, 6H, S$^+$(C$_6$H$_5$)$_3$, J = 7.69 Hz), 7.79-7.71 (br s, 2H, 4-vinylbenzene aromatic), 7.74-7.69 (t, 3H, S$^+$(C$_6$H$_5$)$_3$, J = 7.39 Hz), 7.69-7.61 (t, 6H, S$^+$(C$_6$H$_5$)$_3$, J = 7.64 Hz), 7.10-6.93 (br s, 2H, 4-vinylbenzene aromatic), 4.37-4.03 (br m, overlapping (1H, GMA COOCHH)[5] and (4H, RBMA OCH$_2$CH$_2$O), 3.97-3.78 (br s, 1H, GMA COOCHH),[5] 3.70-3.57 (br m, 8H, RBMA (N(CH$_2$CH$_3$)$_2$)$_2$), 3.27-3.13 (br s, 1H, GMA COOCH$_2$CHOCHH),[5] 2.91-0.14 (br m, aliphatic backbone), 2.86-2.77 (br s, 1H, GMA COOCH$_2$CHOCHH),[5] 2.69-2.57 (br s, 1H, GMA COOCH$_2$CHOCHH),[5] 1.50-1.35 (br m, 9H, TBMA C(CH$_3$)$_3$). GPC: (DMF with 10 mM LiBr, PMMA standards): $M_n$ = 21 kDa, $M_w$ = 46 kDa, Đ = 2.16.

*Characterization of copolymer photoactivity*. Copolymer **1** was dissolved to 10 mg mL$^{-1}$ in MeOH, then drop-cast onto a glass slide heated to 60 °C to afford a polymer film on the slide surface. The film was irradiated (3000 mJ cm$^{-2}$, λ = 365 nm) then placed in a beaker containing a 100 mM HCl solution. Upon contact with the aqueous solution, the colorless film became yellow then colorless as pendent fluorescein moieties were protonated. The film swelled and delaminated from the glass substrate surface within ~2 minutes of contact with the acid solution and remained fully intact in solution for at least 25 hours after delamination. Copolymer **2** was dissolved to 100 mg mL$^{-1}$ in toluene, and drop-cast (5 μL) onto glass slides and allowed to dry without heating. Then, the films were characterized by ATR IR i) without further processing, ii) after heating to 150 °C for 60 s; and iii) after irradiating at λ = 254 nm for a dose of 900 mJ cm$^{-2}$, then heating to 150 °C for 60 s. The change in thickness resulting from cleavage of *t*-butyl esters during photopatterning was quantified by optical profilometry after irradiation (λ = 254 nm) at doses of 12, 25, 50, 100, 200, 450, and 900 mJ cm$^{-2}$ and heating to 150 °C for 60 s.

*Substrate preparation, flow-coating, release, and droplet experiments.* Glass slides (24 x 40 x 0.17 mm$^3$, Fisher Scientific) were cleaned by sonication for 15 minutes each in soapy water, reverse osmosis water, and isopropanol, followed by 15 minutes of surface treatment by UV-ozone to render the surface hydrophilic. Immediately afterwards, a solution of PSS in RO water (20 mg mL$^{-1}$) was applied by spin-coating onto the hydrophilized glass surface (10 s at 500 RPM, then 40 s at 2000 RPM). Samples were partitioned into 2 groups: 1) for experiments with substrate-adhered MS(BC)Ps (Figure 3d, 4, 5, 6c, and S1), PSS-coated slides were laser engraved (2% power, 40% speed, 1000 PPI) at 2-4 mm intervals to afford stripes of bare glass to which MS(BC)Ps would adhere upon flow-coating and release; 2) for experiments with free-floating ribbons (MSDCPs and MSTCPs in Figure 6a-b), the substrate was not laser-engraved. Then, the substrates were fixed to a translating stage, and a razor blade bolted to a stationary mount was lowered to a height of ~ 200 μm above the substrate surface. A polymer-in-toluene solution (5-15 μL of 16 mg mL$^{-1}$ **1** or 4 μL of 4 mg mL$^{-1}$ **2**) was injected between the blade and substrate to afford a capillary bridge 24-36 mm in length. The substrate was translated in 1 mm intervals at 3 mm s$^{-1}$, with a 1.1-1.5 s delay between steps to deposit the MSPs, which were irradiated at i) 3300 mJ cm$^{-2}$ at λ = 365 nm (copolymer **1**) to afford a crosslinked PDMAEMA network, or ii) 200-695 mJ cm$^{-2}$ at λ = 254 nm through a photomask, then heated to 150 °C for 60 s (copolymer **2**), to afford an MSBCP with alternating segments of hydrophobic PTBMA and hydrophilic PMAA. MS(BC)Ps were then cut into 1-4 mm segments *via* laser engraver and subjected to reactive ion etching with O$_2$ plasma for 30 s to remove any residual inter-MS(BC)P polymer film. To release MS(BC)Ps, an aqueous solution was prepared by filling a polystyrene Petri dish (Fisher Scientific, 60 mm diameter, 15 mm depth) with 10 mL of pH buffer solution or RO water. Then, a coated substrate was gently floated on top of the solution and quickly submerged using tweezers. Upon submersion, the underlying PSS layer dissolved to release the MS(BC)Ps. In the case of samples in which the PSS layer was cut before flow-coating, MS(BC)Ps were adhered at one end to the glass surface but were otherwise free to twist, bend, and stretch; for those not subjected to laser cuts, MS(BC)P movement was completely unrestricted. Drawn glass microcapillary tubes were inserted into a Capillary Holder 4 (Eppendorf), which was mounted in a TransferMan 4r micromanipulator (Eppendorf) and connected to a syringe loaded with PFD for injection and withdrawal of the oil phase. Thus equipped, the microcapillary tip was lowered into the aqueous solution to enable hand-controlled manipulation of MS(BC)Ps and droplets. Droplets were introduced by either i) emulsifying a mixture of PFD and the chosen aqueous continuous phase in a 7 mL scintillation vial by ~5 cycles of rapid injection and withdrawal of both liquids (~1 mL aqueous and ~100 μL PFD) through a Pasteur pipette, then quickly injecting the mixture into the Petri dish with released MS(BC)Ps, or ii) directly injecting oil *via* the microcapillary tube.

*Force measurements using a carbon fiber cantilever.* An individual carbon fiber was cut to ~5 mm length, then glued to the end of a capillary tube using Loctite superglue. The cantilever was cut to ~1 mm in length, and the tip dipped into a drop of Loctite superglue then withdrawn to leave a liquid bead attached near the fiber tip. This was cured for a minimum of 12 h, then the capillary tube with affixed cantilever was inserted into a holder, clamped into the hand-controlled micromanipulator, rotated until parallel with the focal plane of the microscope objective, and deflected by bringing it into contact with a glass slide to verify that tip displacement was due exclusively to cantilever deflection. Then, the capillary tube was rotated until the cantilever orientation was out of the objective focal plane and lowered into an aqueous solution reservoir containing MS(BC)Ps and droplets. The superglue bead at the cantilever tip was brought into contact with i) a PFD droplet, then ii) an MS(BC)P that spontaneously wrapped the droplet. The substrate (with attached MS(BC)P end) was translated to load the ribbon-droplet-cantilever assembly and deflect the cantilever, with video data collected at 30 fps. Individual frames were saved in .tif format. Videos were converted to .avi file format using ImageJ image processing software, and the pixel (x,y) positions of key features, including cantilever tip, droplet-cantilever attachment point, MS(BC)P fixed end, and MSBCP inter-segment boundaries were tracked frame-by-frame using Tracker Video Analysis and Modeling Tool. The ribbon vector $\vec{R} = \langle R_x, R_y, 0 \rangle$ was calculated by subtracting the point of ribbon-droplet contact (for MSPs) or an arbitrary inter-segment junction point (for MSBCPs) from the point of cantilever-droplet contact, with assumed 0 z-component because the entire visible ribbon length was within the focal plane. The x- and y-components of the cantilever vector $\vec{C} = \langle C_x, C_y, C_z \rangle$ were calculated by subtracting the position of the cantilever tip from the superglue bead center point, while the z-component was calculated using the Pythagorean theorem $C_x^2 + C_y^2 + C_z^2 = L_{tip}^2$, where $L_{tip}$ is the actual length between bead and tip, measured when the cantilever was parallel to the objective focal plane. The applied force angle $\theta$ was then calculated using the dot product $\vec{R} \cdot \vec{C} = |\vec{R}||\vec{C}| \cos\theta$. Cantilever displacement was measured from the point of cantilever-droplet contact, with 0 deflection defined by the average (x,y) position before MS(BC)P attachment, after MS(BC)P detachment, and/or during periods of slack in the MS(BC)P. The y position was plotted as a function of x position for every frame, and a line of best fit crossing the origin was calculated. The data was then rotated about the origin *via* the rotation matrix with -$\phi$ the angle between the best-fit line and the x axis to afford

$$\begin{bmatrix} \cos\phi & -\sin\phi \\ \sin\phi & \cos\phi \end{bmatrix} \begin{bmatrix} x_0 \\ y_0 \end{bmatrix} = \begin{bmatrix} \delta \\ y_\phi \end{bmatrix},$$

where $\delta$ describes cantilever deflection in the equation

$$F = \frac{3\delta E I}{L^3 \sin\theta}$$

with cantilever modulus $E$ = 230 GPa, moment of inertia for circular cross-section $I = \frac{\pi r_c^4}{4}$, radius $r_c$ = 3.5 μm,[15] and cantilever length $L$ measured as the distance between the cantilever fixed end and the center of droplet attachment (approximated as a point load). In this way, force was calculated for every video frame. Suspended ribbon length ($L_R$, MSPs) was calculated as the distance between the MSP fixed end and the point of ribbon-droplet attachment. Peel length (MSBCPs) was calculated as the distance between the end of the wrapped segment and the point of segment-droplet contact. Moduli of deprotected PMAA MSBCP gel domains were measured from cycle 1 of the same video used for peel force measurements (Video S14) by tracking the (x,y) pixel locations of the segment junction points between PTBMA and PMAA domains; uniaxial swelling ratio was determined by dividing the PMAA segment length (defined as the measured length when cantilever deflection began) by the initial mask feature size (50 μm), while cross-sectional area was determined by multiplying the cross-sectional area (estimated by optical profilometry) by the square of the uniaxial swelling ratio.

Modulus estimate of copolymer 1
*i. Experimental design.* The Young's modulus, $E$, for copolymer **1** MSPs was estimated by examining the deformation of coiled helical MSPs under viscous flow, inspired by a general strategy reported previously.[7] For each selected pH (1, 4, 6, 8, and 10) helical MSPs were subjected to a series of flow steps at increasing flow rate. The helical axial elongation, $H$, was measured as a function of flow velocity (**Figure S4**). We characterized the obtained velocity-extension curves by the slope of the linear regime. The measured slope was combined with an estimated drag coefficient $\xi_{//}$ and several geometrical parameters in a theoretical model to estimate MSP bending modulus $B$ that was then used with measured values of $t$ and $w$ to estimate $E$.[7,8]

*ii. Apparatus.* Helical extension measurements were conducted in PDMS channels (Sylgard 184, DOW Corning) printed using standard soft lithography methods. The channels were coated with a 10% bovine serum albumin (Sigma Aldrich) solution for 15 minutes in order to avoid adsorption on the channel walls. Glass capillaries were similarly coated with a 2% bovine serum albumin solution for 15 minutes. MSPs were released in a pool of the selected buffer solution and displaced using an open glass capillary controlled by a micro-manipulator. The glass capillary was connected to a syringe to catch MSPs by withdrawing and released by expelling liquid. MSPs were captured at one end, then placed in a microfluidic channel connected to the pool. A flow rate $Q$ of the buffer solution was applied to the channel and the resultant helix deformation was tracked by measuring $H$ *via* fluorescence microscopy. The flow velocity $V$ adopted a parabolic distribution in the channel, but as typical helix radii are small compared to the channel size, we estimate a locally uniform flow near the helix. For a given helical MSP, $V$ was taken as the average of the flow field velocity over all the positions occupied by the MSP. The flow field in the channel was computed from the channel dimensions using a derivation from White,[9] and the position of the MSP was measured from captured micrographs.

*iii. Axial elongation measurements.* As seen in Figure S4a, the $H$ does not reach an equilibrium state over the duration of one flow step (usually 30 seconds to 1 minute), verified by immersing helical MSPs in flow for over 1 hour. Moreover, we observed that the helical MSP do not recover the initial length after a flow step (Figure S4a) and that the resting length evolves considerably over the duration of a multi-cycle experiment (Figure S4c). These observations are likely due to creeping of the material under stress induced by the viscous forces. In order to quantify the elastic contribution that is controlled by $E$, we implemented an analysis that decouples the viscous and elastic components of axial extension.

During a single flow step, the deformation has two components: the elastic deformation of the material and the creeping-induced deformation. Assuming constant pulling force and friction, the elastic component is expressed under the form $H_{elastic}(1 - \exp(-t/\tau))$, where $H_{elastic}$ corresponds to the amplitude of the elastic deformation and $\tau$ to the timescale of the helix recovery. We also add a phenomenological term, $\mu t$, where $\mu$ denotes the susceptibility of the material to creeping. The extension curve $H(t)$ is hence fitted by the following semi-phenomenological function: $H(t) - H_0 = H_{elastic}(1 - \exp(-t/\tau)) + \mu t$. $H_0$ is the resting axial length, which is measured and thus not a fitting parameter. Experimentally we find the timescale $\tau$ (typically under 1 s) to be significantly smaller than the typical creeping time $H_0/\mu$ (typically above 100 s). This allows us to clearly separate the elastic regime and the creeping regime. As seen in Figure S4b, agreement with experimental data is good. With this fitting method we recover the elastic extension $\Delta H = H_{elastic} - H_0$ as a function of the flow velocity $V$. The elastic extension, $\Delta H$, as a function of $V$ for 6 different helical MSPs at the same pH is plotted in **Figure S5a**. To characterize the flow-extension curve, we used the heuristic expression proposed by Jawed et al.,[10] based on the simulation of flexibles helices in uniform flow: $\Delta H = \Delta H_{lim}(1 - \exp(-V/V_c))$. Here, the parameter $\Delta H_{lim}$ is the maximum elongation, and the parameter $V_c$ is the characteristic flow speed separating the linear and non-linear regime. As seen in Figure S5a, this expression provides a good description of the helical MSP extension, particularly at low speed. The discrepancies at high speed are likely due to creeping effects. Using this fitting method, we estimated the slope in the small deformation limit as $\Delta H_{lim}/V_c$.

*iv. Modeling.* For a flexible helix immersed in a uniform flow of velocity $V$, the helix elastic axial extension $\Delta H$ can be expressed[7,8] in the small deformation limit as $\Delta H = R^2 L^2 (\xi_{//}/B) V$ where $\xi_{//}$ is the drag coefficient along the tangential direction, $B$ is the bending modulus, $R$ is the helix radius, and $L$ is the total length along the curvilinear abscissa. The MSP cross section is a very shallow triangle with width $w \gg$ thickness $t$. The general form for $B$ of a triangular cross section is $B = (1/36) E w t^3$. $\xi_{//}$ was estimated by approximating the cross section as a rectangle of negligible thickness, giving $\xi_{//} = 4\pi\eta / (2 \ln(8L/w) - 1)$,[11] where $\eta$ is the fluid viscosity. The Young's modulus was calculated as

$$E = \frac{144\,p}{2 \ln\left(\frac{8L}{w} - 1\right)} \eta \frac{R^2 L^2}{w\, t^3} \left(\frac{V_c}{\Delta H_{lim}}\right).$$

*v. MSP width and thickness measurements.* The *t* and *w* in the above expressions correspond to the immersed state of the material. *w* (typically ~ 20 μm) was measured optically *in situ*. However, *t* (typically 100-400 nm) is below the optical resolution limit and was determined by applying a pH-dependent swelling ratio[12,13] to the dry thickness, measured by optical profilometry.

*vi. Results.* The measured values of *E* are presented in Figure S5b. Overall, *E* for copolymer **1** in the immersed state is approximately constant at 100-350 MPa across the pH 1-10 range.

Supplementary Figures

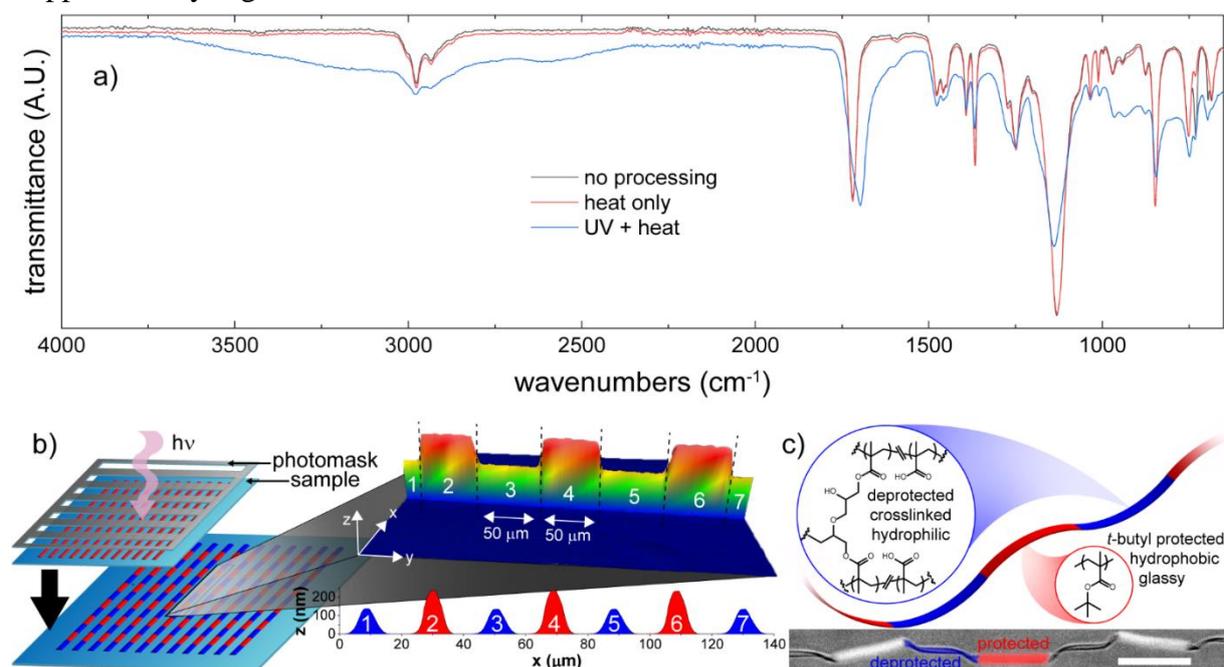

**Figure S1. Copolymer 2 photoactivity and MSBCP characterization.** a) Drop-cast thick films of copolymer **2** were characterized by ATR IR spectroscopy before any treatment (black line), after heating to 150 °C for 60 s (red line), and after irradiating at a dose of 900 mJ cm$^{-2}$ (λ = 254 nm) and then heating to 150 °C for 60 s, revealing carboxylate evolution after irradiation and heating. MSBCPs are prepared by b) irradiating an array of copolymer **2** ribbons through a photomask to afford segments of alternating thickness, shown in 3D (top) and 2D cross-section slices (bottom); c) irradiated domains (blue) are composed of crosslinked poly(methacrylic acid), while masked domains (red) are composed of PTBMA; the false color micrograph represents a typical MSBCP of 50 μm segment length in RO water, with alternating twisted, compliant hydrogel and rigid, brightly fluorescent, hydrophobic domains; scale bar 50 μm.

Photoactivity of copolymer **2** was verified by ATR IR spectroscopy in drop-cast films (**Figure S1a**). A 100 mg mL$^{-1}$ solution of **2** in toluene was drop-cast in 5 μL drops onto glass slides, then characterized i) without further treatment (black spectrum), ii) after heating to 150 °C for 60 s (red spectrum), and iii) after irradiating with λ = 254 nm for a dose of 900 mJ cm$^{-2}$, then heating to 150 °C for 60 s (blue spectrum). The carbonyl peaks were normalized to 20 % absorbance at λ$_{max}$, then converted to % transmittance and offset by 1%. The untreated and heat only samples were identical, with no carboxylic acid -OH signal and a maximum carbonyl signal of 1719 cm$^{-1}$, while a carboxylic acid stretch (3700-2400 cm$^{-1}$) evolved and the carbonyl maximum shifted to 1697 cm$^{-1}$ after irradiation and heating, confirming successful deprotection of *t*-butyl esters. Moreover, ribbons were observed to undergo a change in thickness upon irradiation and heating. In Figure S1b-c, irradiated domains are schematically depicted in blue, while masked domains are shown in red. Optical profilometry (Figure S1b) reveals a thickness loss of up to 0.45x in irradiated segments (labelled 1, 3, 5, and 7) at UV doses of 200 mJ cm$^{-2}$ or larger, while masked domains (labelled 2,4, and 6) retained the original ribbon thickness, consistent with other chemically amplified ribbon and photoresist compositions.[2,14,15] The 3D optical profile data (Figure S1b top) reveals the structure of a typical MSBCP patterned in alternating segments of 50 μm, while the 2D cross section data of each segment (Figure S1b bottom) shows the uniformity in thickness in masked versus irradiated domains. Figure S1c describes a released MSBCP, with schematic structure and inset structure (top) and a micrograph of a typical MSBCP in RO water, including false coloration of a compliant, photobleached, and deprotected hydrogel segment (blue), and a stiff, brightly fluorescent masked segment (red). Crosslinking of pendent epoxides after irradiation and heating was verified by analysis of deprotected domains, which begin to bear load at length ~93 μm in pH 10 buffer solution, an approximate 86% uniaxial strain due to swelling from the original patterned length of 50 μm. This suggests a water volume fraction $\phi_{H2O}$ ~ 0.85 while cyclically bearing the loads required to unwrap an adjacent hydrophobic segment from a PFD droplet.

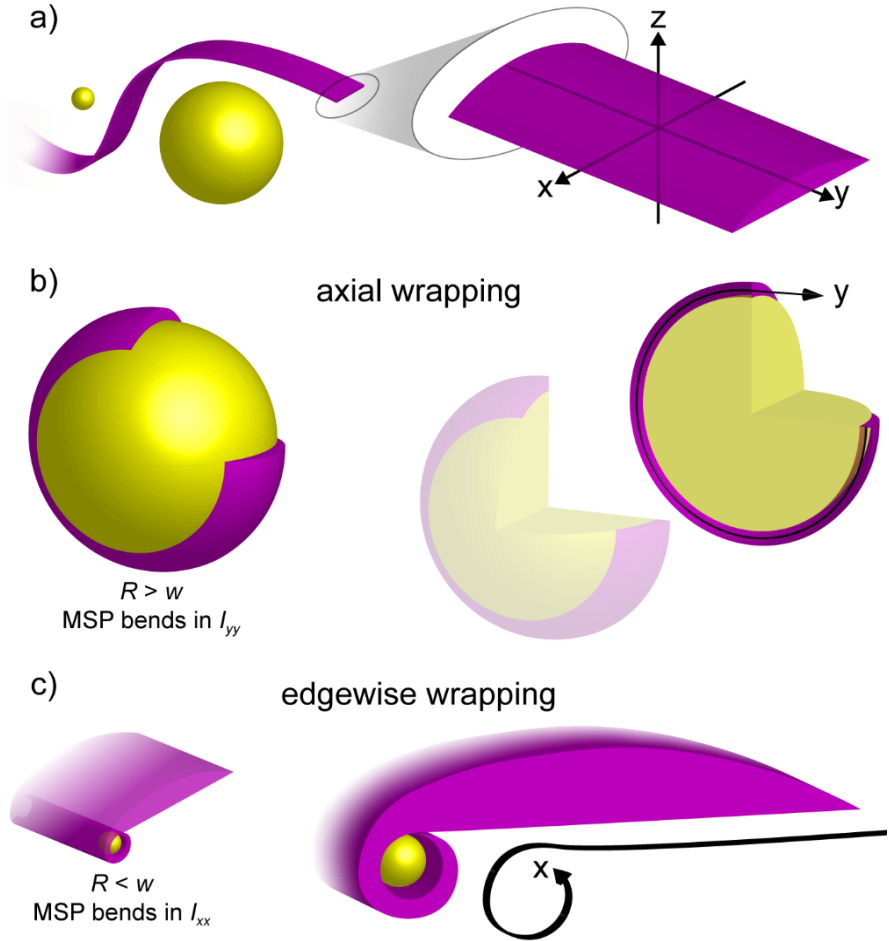

**Figure S2. Axial and edgewise wrapping.** The MSP wrapping axis is expected to depend on the relative size of $R$ and $w$: a) an MSP with a small (left, $R < w$) and large (right $R > w$) droplet. The magnified segment shows the directions of the y- and x-axis relative to the MSP long axis; b) axial wrapping where $R > w$: bending occurs along the y-axis as described by $I_{yy}$; c) edgewise wrapping where $R < w$: bending is anticipated along the x-axis as described by $I_{xx}$, which decreases with $t$ toward the tapered edges of the MSP.

**Figure S2** describes the $R$- and $w$-dependent wrapping modes accessible through moments of inertia $I_{xx}$ and $I_{yy}$. In the case where $R > w$, MSP wrapping depends on the thickness $t$ at the MSP center, and wrapping is observed to proceed along the long ribbon axis, defined as y in Figure S2a. This axial wrapping phenomenon is shown schematically in Figure S2b; because $R > w$, the entire width of the MSP is in contact with the droplet and wrapping proceeds by consuming MSP length and is dependent on bending moment $I_{yy}$. In contrast, we anticipate edgewise wrapping in the case of a droplet with radius $R < w$ (Figure S2c) In this case, MSP bending stiffness becomes vanishingly small toward the MSP edges as $t$ decreases, so droplet contact is predicted to elicit wrapping *via* a rolled-in edge.

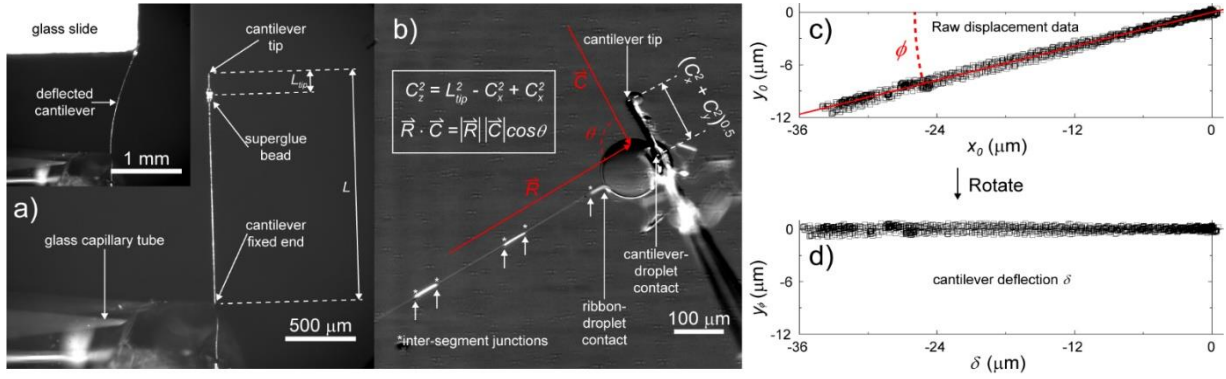

**Figure S3. Cantilever video data acquisition and frame-by-frame processing to determine $\theta$ and $\delta$.** a) The total length $L$ and bead-to-tip length $L_{tip}$ were imaged and measured, then the cantilever was deflected (inset) using a glass slide on a translating; b) video data of an MSBCP-droplet-cantilever system under applied load. (x,y) pixel locations of key features, including cantilever tip, cantilever-droplet contact point, ribbon-droplet contact point, and MSBCP inter-segment junctions were tracked frame-by-frame. The force angle $\theta$ was calculated *via* the dot product of the cantilever and ribbon vectors $\vec{R}$ and $\vec{C}$ in each frame, while cantilever deflection $\delta$ was determined from the raw (x,y) displacement data of the cantilever-droplet contact point by rotating about the origin to lie on the x-axis.

Before measurement of cantilever deflection, a test deflection was carried out at low magnification (**Figure S3a**) by bringing into contact with a glass slide on a translating stage (Figure S3a inset). Frames from this experiment were used to ensure that the cantilever fixed end remained stationary during deflection and to measure the full cantilever length $L$ and the distance from the superglue bead to the cantilever tip $L_{tip}$. Accurate force measurement required quantification of cantilever deflection $\delta$ and applied force angle $\theta$. The ribbon vector $\vec{R}$ (Figure S3b) was assumed to have negligible z-component because the ribbon was in the focal plane of the lens, while the x- and y- components were tracked frame-by-frame (see **Methods**). The cantilever vector $\vec{C}$ had a significant z-component that was determined using $C_x^2 + C_y^2 + C_z^2 = L_{tip}^2$, where $C_y^2 + C_z^2$ was determined visually frame by tracking the bead center and cantilever tip, and $L_{tip}$ was a constant as measured in Figure S3a. $\theta$ was calculated *via* the dot product of $\vec{R}$ and $\vec{C}$ (see **Methods**). Similarly, cantilever deflection was determined by tracking the (x,y) pixel location of the point of cantilever-droplet contact (in the case of Figure 6, Figure S3, and Video S14, this was taken to be the center of the superglue bead) against an origin defined by the average position in the absence of load. This raw data was converted to microns (Figure S3c top), then rotated about the origin such that the line of best fit was y = 0; $\delta$ was defined for each frame as the rotated x displacement data (Figure S3c bottom).

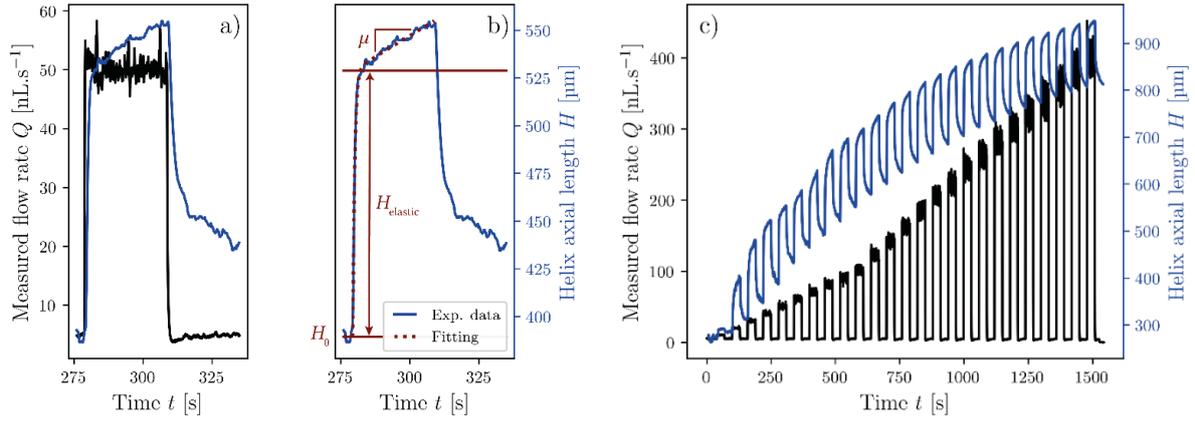

**Figure S4. Helix extension under axial flow.** a) Typical applied flow step and following relaxation: the measured buffer solution flow rate (black) and measured helical axial length (blue) are plotted as a function of time. The flow is not completely stopped during the relaxation phase, keeping instead a vanishing value $Q = 2$ nL/s. The syringe pump responds more quickly when changing the flow rate than when starting the flow. The viscosity of the buffer solution is always 1.0 mPa.s. For all experiments, the channel width is 250 μm and height is 650 μm. b) Fitting of the previous helix extension curve using a semi phenomenological function (3 fitting parameters). c) Typical full flow cycle applied; Figure S4a is extracted from this curve.

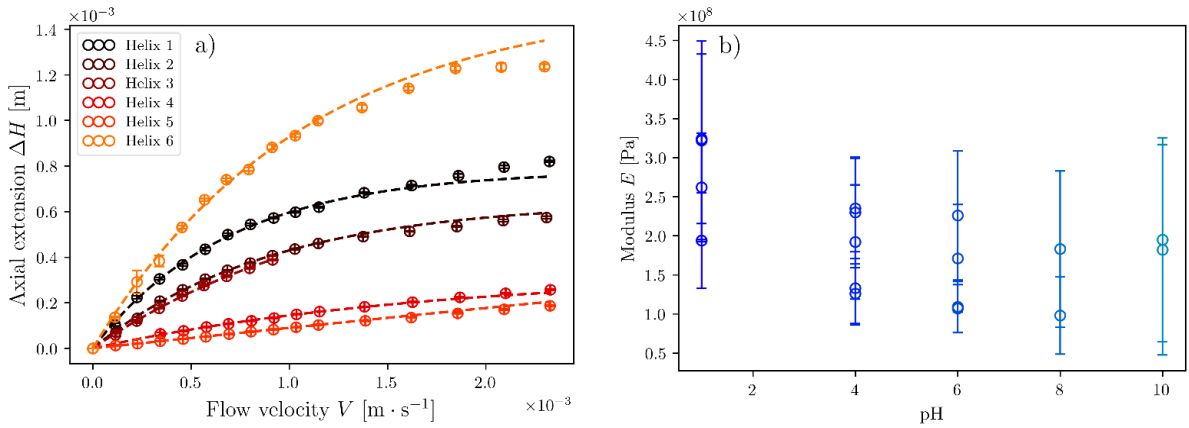

**Figure S5.** a) Flow-extension curves for 6 different copolymer **1** helical MSPs immersed in a pH 4 buffer solution with heuristic fitting. b) Measured values for the Young's modulus $E$ across the 1-10 pH range.

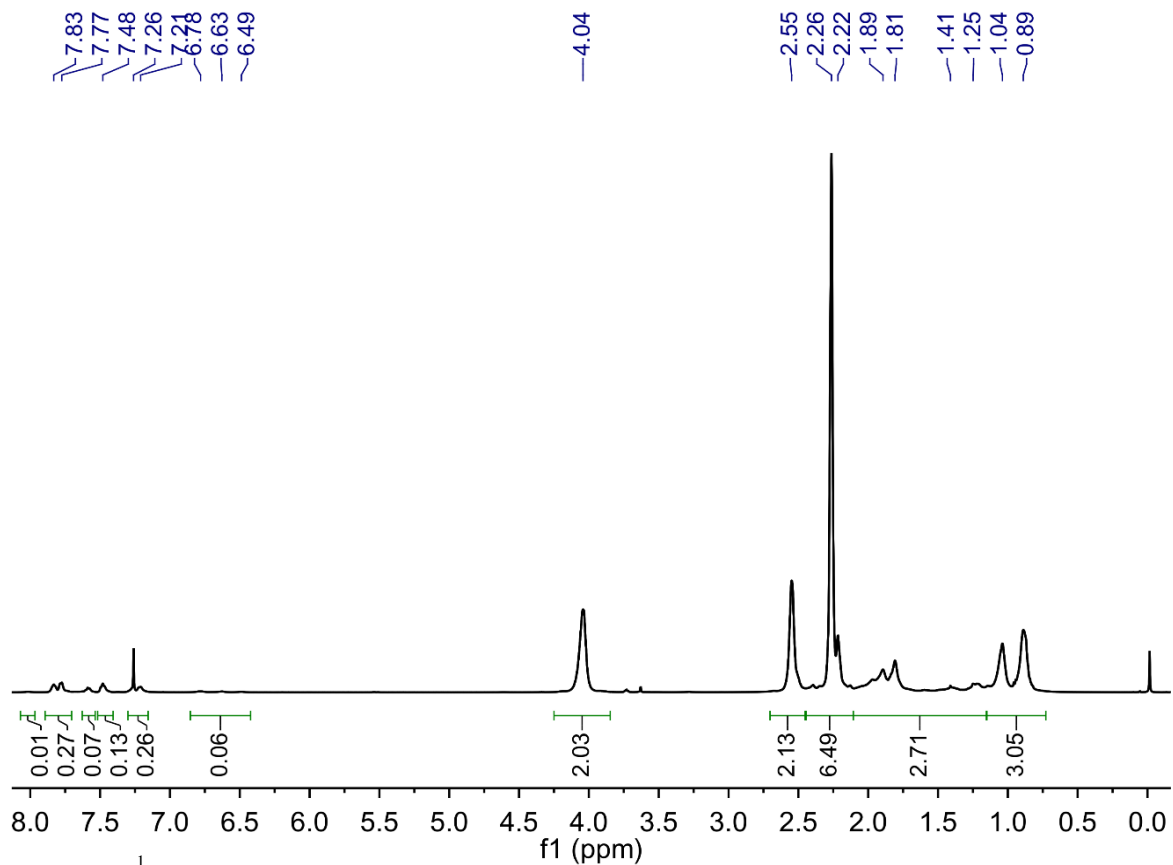

**Figure S6.** $^1$H NMR spectrum of copolymer **1**

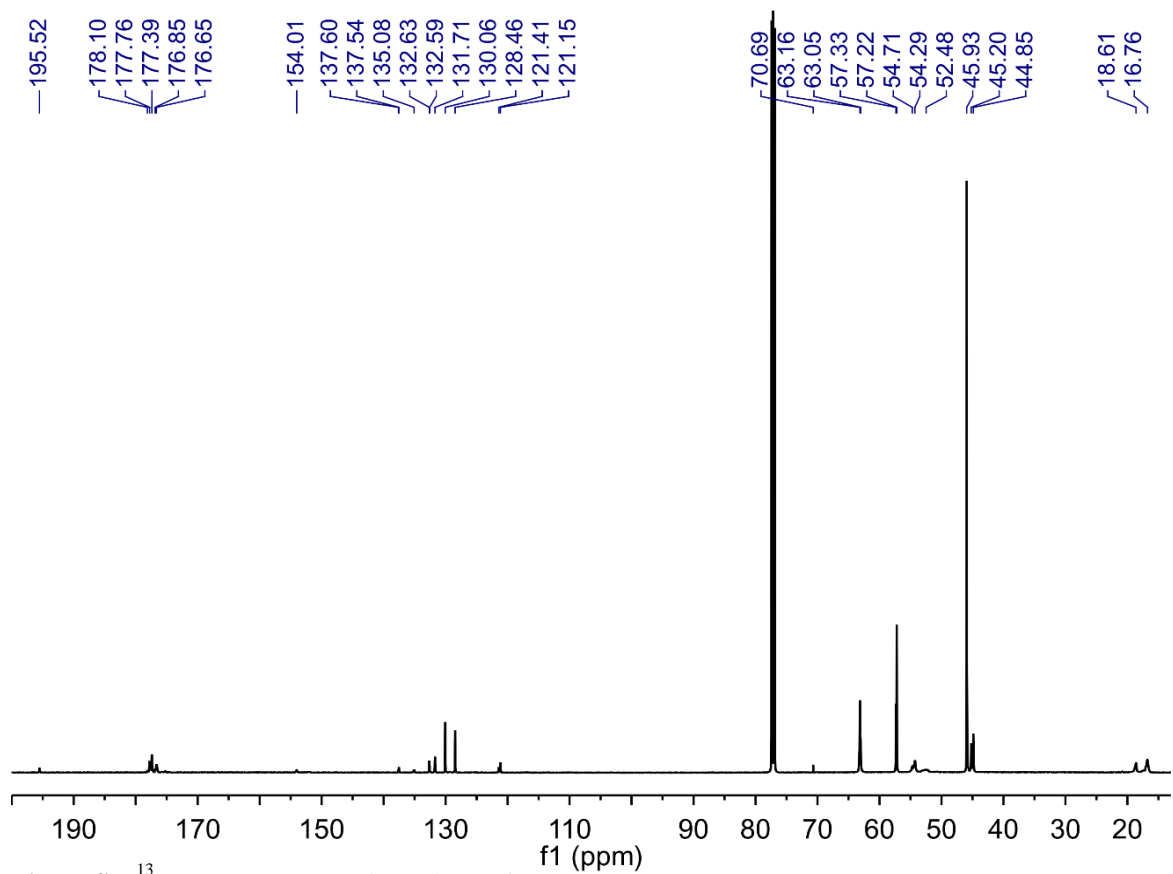

**Figure S7.** $^{13}$C NMR spectrum of copolymer **1**

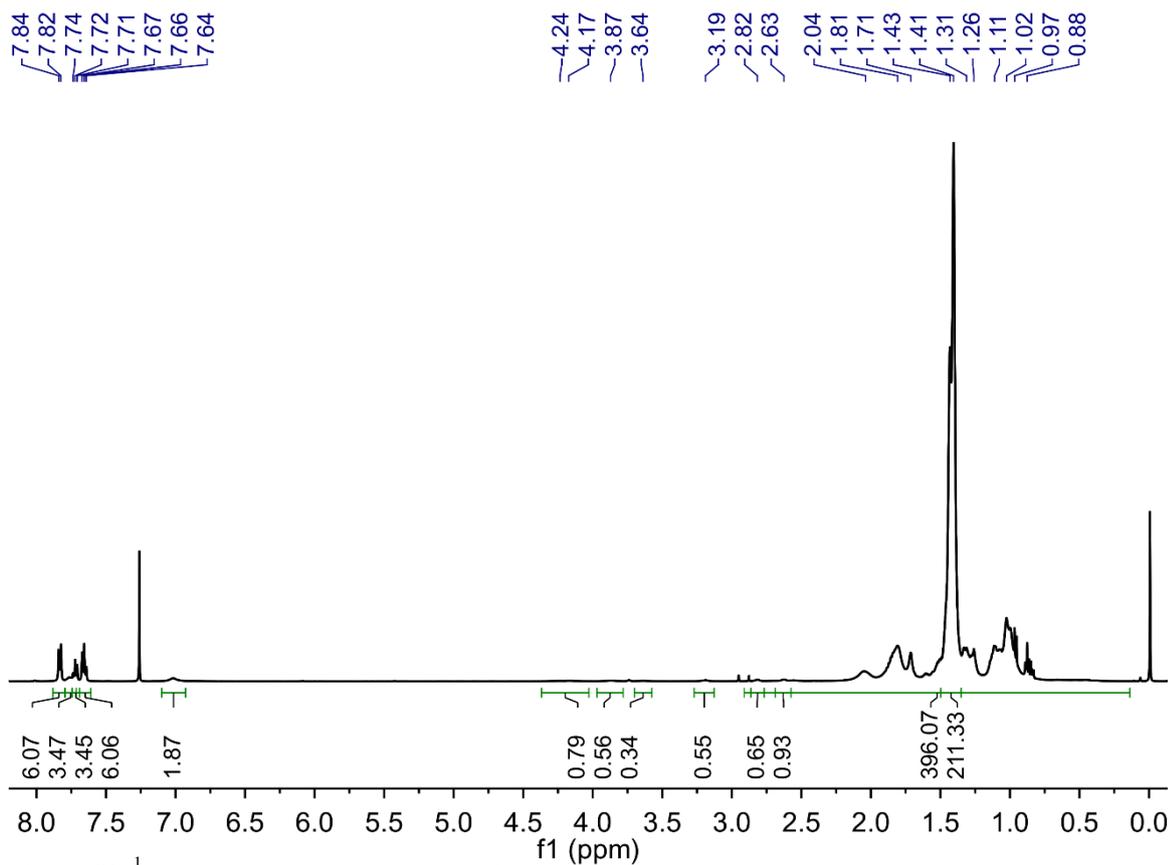

**Figure S8.** $^1$H NMR spectrum of copolymer **2**

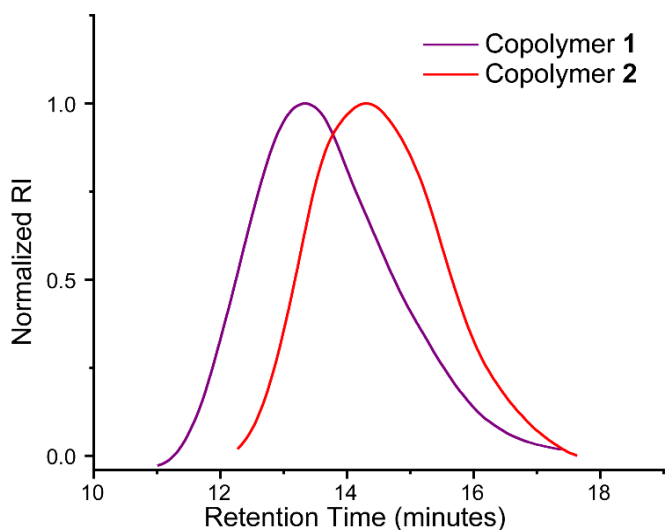

**Figure** S8. GPC traces of copolymers **1** and **2**.

Supplementary Videos

**Video S1.** A helical MSP (copolymer **1**, helix radius $r = 38$ μm) in pH 1 buffer solution with one end fixed to the substrate surface in contact with a PFD droplet ($R = 132$ μm). As the substrate with adhered MSP end is translated to the left, the coiled helical MSP stretches until it detaches from the droplet surface and recoils through the solution.

**Video S2.** A helical MSP (copolymer **1**, helix radius $r = 55$ μm) in pH 4 buffer solution with one end fixed to the substrate surface in contact with a PFD droplet ($R = 335$ μm). As the substrate with adhered MSP end is translated to the left, the coiled helical MSP stretches until 4 coils detach from the droplet surface (time $T \sim 2.3$ s). Upon further stretching, the droplet is pulled from the microcapillary tip by the adhered MSP spring.

**Video S3.** A helical MSP (copolymer **1**) in pH 6 buffer solution. The left end of the helix is attached to the substrate, while the right end became fixed to the substrate after release, affording a structure with 2 fixed ends. As a PFD droplet is brought into contact with the helical ribbon, the two bodies slide past each other without apparent adhesion.

**Video S4.** A short MSP segment (copolymer **1**, length ~ 400 μm) in pH 8 buffer solution is adhered at one end to the surface of a droplet and at the far end to the substrate. Ribbon and droplet are manipulated through the solution *via* microcapillary tip and translating stage, revealing selective adhesion at the ribbon tip.

**Video S5.** MSPs (copolymer **1**) in pH 10 buffer solution wrapped around a droplet. The droplet is anchored in place by the fixed end of a wrapped ribbon, while the microcapillary tube and translating stage are used to "unwrap" the droplet.

**Video S6.** An MSP (copolymer **1**) is held in tension by the microcapillary tip to control wrapping in pH 10 buffer solution. As slack is added to the system by bringing the MSP end toward the wrapped droplet, the MSP continues to wrap until it overlaps an existing coil, arresting the wrapping event.

**Video S7.** A droplet is inflated next to an MSP (copolymer **1**) in pH 10 buffer solution. To the left (out of frame), the MSP is fixed to the substrate surface; to the right it floats freely. When the droplet touches the MSP, spontaneous wrapping occurs until a defect in the ribbon causes self-overlap, stopping the wrapping event before the ribbon length is consumed and creating a droplet with a pendent arm. To the left, wrapping continues until the ribbon is pulled tight against the substrate-adhered end.

**Video S8.** A droplet is inflated until it comes into contact with an MSP (copolymer **1**) in pH 10 buffer solution. The ribbon is fixed to the substrate to the left (out of frame) and floats freely to the right. Upon contact, the ribbon spontaneously wraps the droplet until the free end is consumed and the ribbon is pulled tight against the substrate-bound end to the left, final droplet radius $R = 360$ μm.

**Video S9.** An MSP (copolymer **1**) in pH 10 buffer with one end adhered to the substrate surface (left, out of frame) is partially wrapped around a droplet ($R = 88$ μm) that is adhered to a superglue bead near the end of a carbon fiber cantilever. The ribbon-droplet and cantilever-droplet interfaces are loaded by translating the substrate to the left to pull on the ribbon. Cantilever deflection is used to quantify the applied loads as the system is loaded, unloaded, and then loaded until detachment of the ribbon from the droplet surface.

**Video S10.** Copolymer **2** MSBCP with 500 μm block length has selectively wrapped a droplet ($R$ = 110 μm) in pH 10 buffer solution to afford a droplet with a single arm extended into solution.

**Video S11.** An MSBCP in pH 10 buffer solution attached to the same droplet as in **Video S11** *via* selective wrapping to add a second arm.

**Video S12.** More MSBCPs in pH 10 buffer solution adhered to and selectively wrapped around the same droplet as in Videos S10 and S11. In this case, we observe mixed assembly modes, including end-on adhesion, adhesion of the hydrophilic segment without wrapping ($R < R_c$), and selective segmental wrapping of the hydrophobic block ($R > R_c$).

**Video S13.** MSBCPs are picked up from the substrate using a PFD droplet ($R$ = 150 μm) adhered to the microcapillary tip in 500 mM NaOH solution.

**Video S14.** Cantilever deflection of an MSBCP (patterned segment length 50 μm) with one end adhered to the substrate surface, and a far segment adhered to a cantilever-bound droplet ($R$ = 60 μm). The system is twice subjected to a full cycle of loading until peel initiation and unloading until re-wrap. On the third cycle, peel is initiated, then propagated until complete detachment of the adhered segment.

References for Supplementary Information